\documentclass[prd,onecolumn,nofootinbib,amsfonts,amssymb,amsmath,12pt,preprint]{revtex4}
\usepackage{latexsym,graphicx,psfrag,here,bm}
\newcommand{\nn}{\nonumber}
\newcommand{\Eref}[1]{Eq.~(\ref{#1})}

\newcommand{\Fref}[1]{Fig.~\ref{#1}}

\newcommand{\eqnspc}{\;}

\newcommand{\ordo}[1]{\ensuremath{{\cal O}(#1)}}

\renewcommand{\Re}{\text{Re}}
\renewcommand{\Im}{\text{Im}}

\newcommand{\lamt}{V\hspace{-0.2em}\raisebox{-0.35ex}{$\scriptstyle
    ts$}^\star V_{tb}}

\newcommand{\alphas}{\ensuremath{\alpha_{\text{s}}}}
\newcommand{\as}{\ensuremath{\alpha_{\text{s}}}}

\newcommand{\alphasFourPi}{\ensuremath{\frac{\alphas}{4 \pi}}}

\newcommand{\bea}{\begin{eqnarray}}
\newcommand{\eea}{\end{eqnarray}}
\newcommand{\beq}{\begin{equation}}
\newcommand{\eeq}{\end{equation}}

\newcommand{\mub}{\ln\left(\frac{m_b}{\m}\right)}
\newcommand{\mubb}{\ln^2\left(\frac{m_b}{\m}\right)}

\def\bra{\langle}
\def\ket{\rangle}

\def\qr{q \cdot r}

\def\a{\alpha}
\def\b{\beta}
\def\g{\gamma}
\def\d{\delta}
\def\e{\epsilon}
\def\p{\pi}

\def\l{\lambda}
\def\m{\mu}

\newcommand{\dslash}[1]{\mbox{$\not \! #1$}}

\begin{document}


\title{Fermionic NNLL corrections to $\bm{b \rightarrow s \gamma}$}
\author{Kay Bieri}
\email{bierik@itp.unibe.ch}
\author{Christoph Greub}
\email{greub@itp.unibe.ch}
\affiliation{Institut f\"ur Theoretische Physik, Universit\"at Bern,
  CH-3012 Bern, Switzerland}
\author{Matthias Steinhauser}
\affiliation{\mbox{II}. Institut f\"ur Theoretische Physik,
  Universit\"at Hamburg, D-22761 Hamburg, Germany}
\begin{abstract}
  In this paper we take the first step towards a complete
  next-to-next-to-leading logarithmic (NNLL) calculation of the
  inclusive decay rate for $B \to X_s\gamma$.
  We consider the virtual corrections of
  order $\alphas^2 n_f$ to the matrix elements of the
  operators ${O}_1$, ${O}_2$ and ${O}_8$
  and evaluate the real and virtual contributions to ${O}_7$.
  These corrections are expected to be numerically important. 
  We observe a strong cancelation between the contributions from the 
  current-current operators and $O_7$ and 
  obtain, 
  after applying naive non-abelianization, a reduction of the branching ratio
  of $3.9\%$ (for $\mu=3.0$~GeV) and an increase of
  $3.4\%$ (for $\mu=9.6$~GeV).
\end{abstract}
\preprint{BUTP-2003/03}
\preprint{DESY 03-013}
\maketitle

\section{Introduction}

Currently, measurements of the
inclusive branching ratio BR$(B \to X_s \gamma)$ are provided
by CLEO~\cite{Chen:2001fj} (Cornell), by the
$B$ factory Belle~\cite{Abe:2001hk} (KEK),  by
ALEPH~\cite{Barate:1998vz} (CERN) and by the preliminary
BABAR~\cite{Aubert1,Aubert2} (SLAC) results,
leading to a world average of~\cite{Jessop}
\begin{eqnarray}
  {\rm BR}(B \to X_s \gamma)_{\rm exp}
  = (3.34 \pm 0.38) \times 10^{-4}
  \, .
  \label{eq:bsgexp}
\end{eqnarray}
This experimental average is
in good agreement with the theoretical prediction based on the
Standard Model (SM) 
including next-to-leading logarithmic (NLL) QCD corrections
supplemented by certain classes of leading order electroweak 
terms~\cite{Czarnecki:1998tn,Kagan:1999ym,Baranowski:1999tq,GH00}.
For a recent status report on inclusive rare $B$ decays
and a complete list of references on NLL calculations 
of $\rm{BR}(B \to X_s \gamma)$ 
the reader is referred to \cite{Hurth:2003vb}.
In earlier analyses~\cite{Chetyrkin:1996vx,Kagan:1999ym,Ciuchini:1997xe,BG98,
Buras:1997},
the ratio $m_c/m_b$, which enters the calculation
of the decay width $\Gamma(B \to X_s \gamma)$ for the first time 
at the NLL level,
was tacitly interpreted to be the ratio of the pole quark masses. Using
$m_c/m_b=0.29 \pm 0.02$, one obtains
${\rm BR} (B \to X_s \gamma)_{\rm SM}= (3.35 \pm 0.30)\times 10^{-4}$,
where the errors due to the uncertainties in the various input parameters
and the estimated uncertainties due to the left-over renormalization
scale dependence were added in quadrature. More recently, Gambino
and Misiak~\cite{Gambino:2001ew} pointed out that the branching
ratio rises to ${\rm BR}(B \to X_s \gamma)_{\rm
SM}= (3.73 \pm 0.30)\times 10^{-4}$~\cite{Gambino:2001ew}
(see also~\cite{Buras:2002tp}),
if one interprets $m_c/m_b$ to be
$\overline{m}_c(\mu)/m_b=0.22 \pm
0.04$, where $\overline{m}_c(\mu)$ is the charm quark mass in
the $\overline{\rm MS}$-scheme, evaluated at a scale $\mu$ in the range
$m_c < \mu < m_b$, and $m_b$ is the bottom quark $1S$ mass.

Despite the current
theoretical dispersion on the branching ratio, the agreement between
the present experimental results
and the SM is quite impressive and this has been used to derive
model independent bounds on the Wilson coefficients $C_7(m_W)$ and
$C_8(m_W)$ (see, for example, Ref.~\cite{Ali:2002jg}).

Formally, the approximately $11\%$ discrepancy in the branching ratio,
stemming from the two different schemes for $m_c/m_b$, is a NNLL effect.  As
the measurements of ${\rm BR}(B \to X_s \gamma)$ will become much more precise
in the near future, it will become mandatory to systematically extend the
theoretical predictions to NNLL precision, in order to fully exploit this
process in the search for new physics.

To illustrate the complexity of such a calculation, we
briefly explain the theoretical framework. Usually, one works in
the effective field theory formalism of the SM, where the $W$ boson and
heavier degrees of freedom are integrated out.
This results in an effective Hamiltonian in which operators
up to dimension six are retained.
Adopting the operator definition of~\cite{Chetyrkin:1996vx},
the relevant Hamiltonian to describe the processes
$b \to s \gamma$, $b \to s g$ and $b\to s\gamma g$ reads
\begin{eqnarray}
  H_{\rm eff} &=& -\frac{4G_F}{\sqrt{2}} \lambda_t \sum_{i=1}^{8}
  C_i(\mu) O_i(\mu)
  \,,
\end{eqnarray}
where $G_F$ is the Fermi coupling constant,
$\lambda_t=\lamt$ (with
$V_{ij}$ being elements of the Cabibbo-Kobayashi-Maskawa matrix) and
$C_i(\mu)$ are the Wilson coefficient functions evaluated at the scale $\mu$.
For practical reasons it is more convenient 
to use instead of the original functions $C_i(\mu)$
certain linear combinations,
the so--called ``effective Wilson coefficients''
$C_i^{\,{\rm eff}}(\mu)$ introduced in~\cite{Buras:xp,Chetyrkin:1996vx}:
\begin{eqnarray}
  \label{cieff}
  C_{i}^{\,{\rm eff}}(\mu) &=& C_i(\mu) \,, \qquad (i=1,\ldots,6) \,, 
  \nn \\
  C_7^{\,{\rm eff}}(\mu) &=& C_7(\mu) +\sum_{i=1}^6 y_i C_i(\mu) \,,
  \nonumber \\
  C_8^{\,{\rm eff}}(\mu) &=&   C_8(\mu) +\sum_{i=1}^6 z_i C_i(\mu) \,,
\end{eqnarray}
where $y_i$ and $z_i$ are defined in such a way
that the leading order matrix elements
$\langle s \gamma |{O}_i|b \rangle$ and
$\langle s g      |{O}_i|b \rangle$ ($i=1,\ldots,6$) are
absorbed in the leading order terms
of $C_7^{\,{\rm eff}}(\mu)$ and $C_8^{\,{\rm eff}}(\mu)$.
The explicit values of $\{y_i\}$ and $\{z_i\}$,
$y=(0,0,-\frac{1}{3},-\frac{4}{9},-\frac{20}{3},-\frac{80}{9})$,
$z=(0,0,1,-\frac{1}{6},20,-\frac{10}{3})$
were obtained in Ref.~\cite{Chetyrkin:1996vx}
in the $\overline{\rm MS}$ scheme using
fully anticommuting $\gamma_5$ 
which is also adopted in the present paper.

The operators relevant for our calculation read
\begin{eqnarray}
O_1 &=&
 (\bar{s}_L \gamma_\mu T^a c_L) \,
 (\bar{c}_L \gamma^\mu T^a b_L) \, , \nonumber \\
O_2 &=&
 (\bar{s}_L \gamma_\mu c_L)\,
 (\bar{c}_L \gamma^\mu b_L)\, , \nonumber \\
O_4 &=&
 (\bar{s}_L \gamma_\mu T^a b_L)
 \sum_q
 (\bar{q} \gamma^\mu T^a q)\, ,  \nonumber \\
O_7 &=&
  \frac{e}{16\pi^2} \, \overline{m}_b(\mu) \,
 (\bar{s}_L \sigma^{\mu\nu} b_R) \, F_{\mu\nu} \, , \nonumber \\
O_8 &=&
  \frac{g_s}{16\pi^2} \,
\overline{m}_b(\mu) \,
 (\bar{s}_L \sigma^{\mu\nu} T^a b_R)
     \, G^a_{\mu\nu}\, .
\label{opbasis}
\end{eqnarray}
Here $e=\sqrt{4\pi\alpha_{\rm em}}$ and $g_s=\sqrt{4\pi\alpha_s}$
denote the electromagnetic and strong coupling constants,
respectively.
Furthermore, $F_{\mu\nu}$ and $G_{\mu\nu}^a$ are the corresponding
field strength tensors and
$L=(1-\gamma_5)/2$ and $R=(1+\gamma_5)/2$ stand for left- and
right-handed projection operators. The factor
$\overline{m}_b(\mu)$ in the definition of $O_7$ and $O_8$
denotes the bottom mass in the $\overline{\rm MS}$ scheme.

For a complete NNLL calculation in this framework, the evaluation
of three parts is necessary: (1) the computation of the
matching coefficients to order $\alpha_s^2$ which requires a three-loop
calculation;
(2) the evaluation of the anomalous dimension matrix to order
$\alpha_s^3$ where four-loop diagrams are involved; and (3)
the calculation of the order $\alpha_s^2$ QCD
corrections to the matrix elements $\langle s \gamma|O_i(\mu)|b \rangle$
($\mu$ is of order $m_b$) which, depending on
the operator, is either a two- or three-loop calculation.

The relatively large dependence of the NLL prediction for
${\rm BR} (B \to X_s \gamma)_{\rm SM}$ on the scheme
for $m_c/m_b$ illustrates that NNLL effects,
in particular those related to step (3), can be rather large.

At this point we should stress, that the issue
related to the definition of $m_c/m_b$ serves us as a
motivation to initiate a NNLL calculation for
${\rm BR}(B \to X_s \gamma)$. In the present paper
we are working out a class of NNLL corrections
(to be specified below)
to step (3), which is not related
to the $m_c/m_b$ issue. However, in many other processes
it is known that the kind of terms considered in this paper
are the source of very important higher order corrections.

In this paper we consider those corrections of order $\alpha_s^2$ to the
matrix elements for $B \to X_s \gamma$ associated with
the operators $O_1$, $O_2$, $O_7$ and $O_8$ which involve
a closed fermion loop. It is needless to say, that
at the same time also 
the matching coefficients and the anomalous dimension matrix
should be improved accordingly.
Motivated by the fact that the NLL corrections to the
matrix elements were numerically more important than the
improvements in the Wilson coefficients, we assume for the time
being that this could also be the case at the NNLL level.
Therefore, we only concentrate on NNLL corrections to the matrix elements.
In principle also the contributions from the operators
${O}_i$ $(i=3,\ldots,6)$ should be considered. However, as the
corresponding Wilson coefficients are small, we neglect these
contributions.
Furthermore, we also neglect the NNLL bremsstrahlung corrections
to the interferences
$(O_1,O_1)$,
$(O_1,O_2)$,
$(O_2,O_2)$,
$(O_1,O_7)$,
$(O_1,O_8)$,
$(O_2,O_7)$,
$(O_2,O_8)$,
$(O_7,O_8)$ and
$(O_8,O_8)$,
since these terms are infrared finite for vanishing gluon energy 
and numerically relatively small at the NLL level~\cite{Greub:1996tg}.

The fermionic corrections we are interested in are essentially generated by
inserting a one-loop fermion bubble into the gluon propagator of the lower
order Feynman diagrams. For the numerical evaluation we will assume that
$n_f=5$ massless fermions are present in the fermion
loop.

Once the corrections of ${\cal O}(\alpha_s^2 n_f)$ are available, it is
suggestive to use the
hypothesis of naive non-abelianization (NNA)~\cite{Beneke:1994qe} in order to
estimate the complete corrections of order $\alpha_s^2$.
This is based on the observation
that the lowest coefficient of the QCD $\beta$ function,
$\beta_0=11-2n_f/3$, is quite large and thus it is expected
that the replacement of $n_f$ by $-3\beta_0/2$ may lead to a good
approximation of the full order $\alpha_s^2$ corrections.
There are many physical observables, where NNA provides an excellent
approximation to the full two-loop 
corrections~\cite{Brodsky:1982gc,Luke:1994yc}
like the inclusive cross section $e^+e^-\to\mbox{hadrons}$, the hadronic
$\tau$ decay or the two-loop relation between the $\overline{\rm MS}$ and pole
quark mass. In particular, we want to mention the semileptonic decay 
$\Gamma(b\to c l\nu_l)$ where the deviation of the 
$\alpha_s^2\beta_0$ terms from the complete $\alpha_s^2$
result~\cite{Czarnecki:1998kt} is less than 20\%.
We also note that the ${\cal O}(\as^2 \beta_0)$ corrections to
the photon energy spectrum in $B \to X_s \gamma$
(away from the endpoint) were calculated
in Ref.~\cite{Ligeti:1999ea}.

Our presentation is organized as follows: in Section~\ref{sec:O2} we
discuss the virtual corrections associated with ${O}_{1,2}$ and
compute in Section~\ref{sec:o7calc} both the real and virtual corrections
to ${O}_7$. The virtual corrections to ${O}_8$ are
considered in Section~\ref{sec:o8calc}.
In Section~\ref{sec:num} we combine our findings with the existing
NLL results and perform a numerical analysis showing the importance of
our new terms.
Finally, Section~\ref{sec:concl} contains our conclusions.
In the appendix supplementary material is provided:
Appendix~\ref{sec:build} contains the building blocks which are
useful for the practical calculations and in
Appendix~\ref{sec:threeloopresults}
detailed analytical results are presented for
the corrections to the matrix element $\bra s\gamma|O_2|b\ket$.
For completeness the results of the order $\alpha_s$ corrections are
listed in Appendix~\ref{sec:rili} and intermediate results needed for 
the matrix element $\bra s\gamma|O_7|b\ket$ are given in
Appendix~\ref{sec:zfactors}.
In Appendix~\ref{sec:cutoff} the results are provided which are necessary to 
discuss the branching ratio 
${\rm BR}(b \to X_s \gamma)_{E_\gamma\ge E_{\rm cut}}$
where $E_{\rm cut}$ represents a cut-off on the photon energy.

\section{Virtual corrections to $\bm{b \to s \g}$ associated with 
$\bm{O_1}$ and $\bm{O_2}$}
\label{sec:O2}
In this section we derive the (renormalized) order $\alpha_s^2$
corrections to the matrix elements 
$ \langle s \gamma |O_1| b \rangle $ and 
$\langle s \gamma |O_2| b \rangle $. 
Thereby only the contributions 
proportional to the number of fermion flavors, $n_f$, are taken into account.
We show at the end of this section that the result for 
$\langle s \gamma |O_1| b \rangle$ can easily be obtained from the one for  
$\langle s \gamma |O_2| b \rangle$. Therefore, we concentrate in the following
on the calculation of  
the renormalized matrix element $M_2$,
\begin{equation}
  M_2 = \bra s \gamma | O_2 | b \ket\,,
\end{equation}
which is conveniently written in the form
\begin{eqnarray}
  M_2 = M_2^{(0)} + M_2^{(1)} + M_2^{(2)}
  \label{eq:M2def}
  \,.
\end{eqnarray}
The superscript counts the factors of $\alpha_s$.
The leading term vanishes, i.e. $M_2^{(0)}=0$ and the ${\cal O}(\as)$
calculation has been performed in~\cite{Greub:1996tg}.
In the following, we discuss the ${\cal O}(\alpha_s^2 n_f)$ term,
$M_2^{(2)}$.
In Subsection~A we present the calculation and results of the
dimensionally regularized three-loop diagrams, while Subsection~B
is devoted to the calculation of the counterterms. In 
Subsection~C 
we combine the results of the three-loop results with the counterterms and
derive the renormalized expression $M_2^{(2)}$. 

\subsection{Regularized three-loop corrections to 
$\bm{\langle s \gamma|O_2|b \rangle$}}
\label{sec:threeloop}

\begin{figure}[t]
  \includegraphics[height=7.9cm]{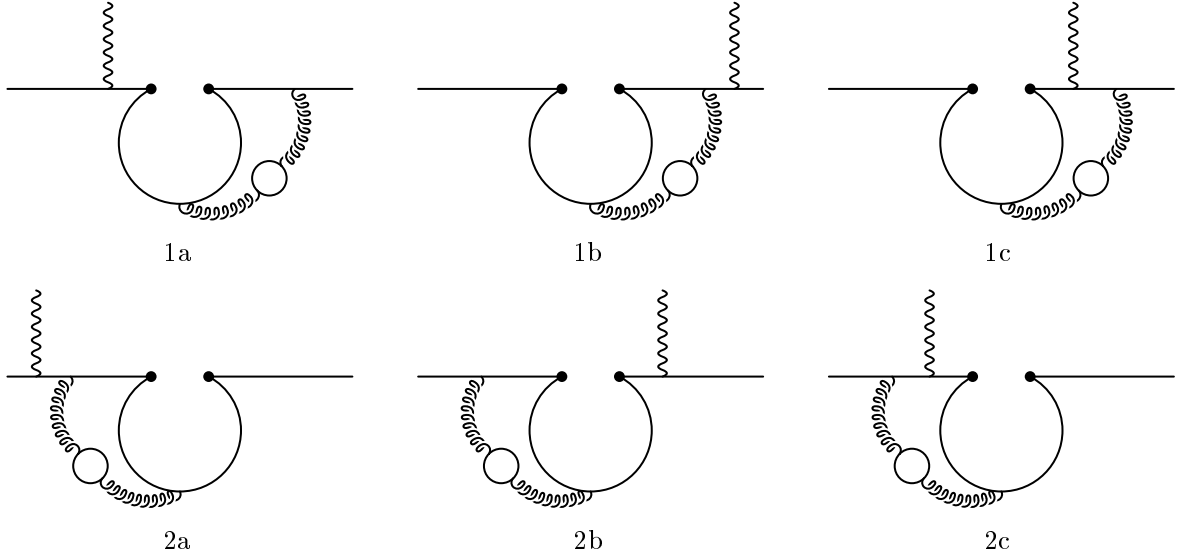}
  \caption{\label{fig:M1M2} Diagrams 1a--c and 2a--c associated with the
    operator $O_2$. The photon is represented by a wavy line and is emitted
    from a down-type quark in all the diagrams. The virtual gluons are
    represented by curly lines. The sum of the first three graphs is denoted
    with $M_{2,{\rm bare}}^{(2)}(1)$,
    whereas the sum of the second three diagrams is
    called $M_{2,{\rm bare}}^{(2)}(2)$.}
\end{figure}

\begin{figure}[t]
  \includegraphics[height=6.3cm]{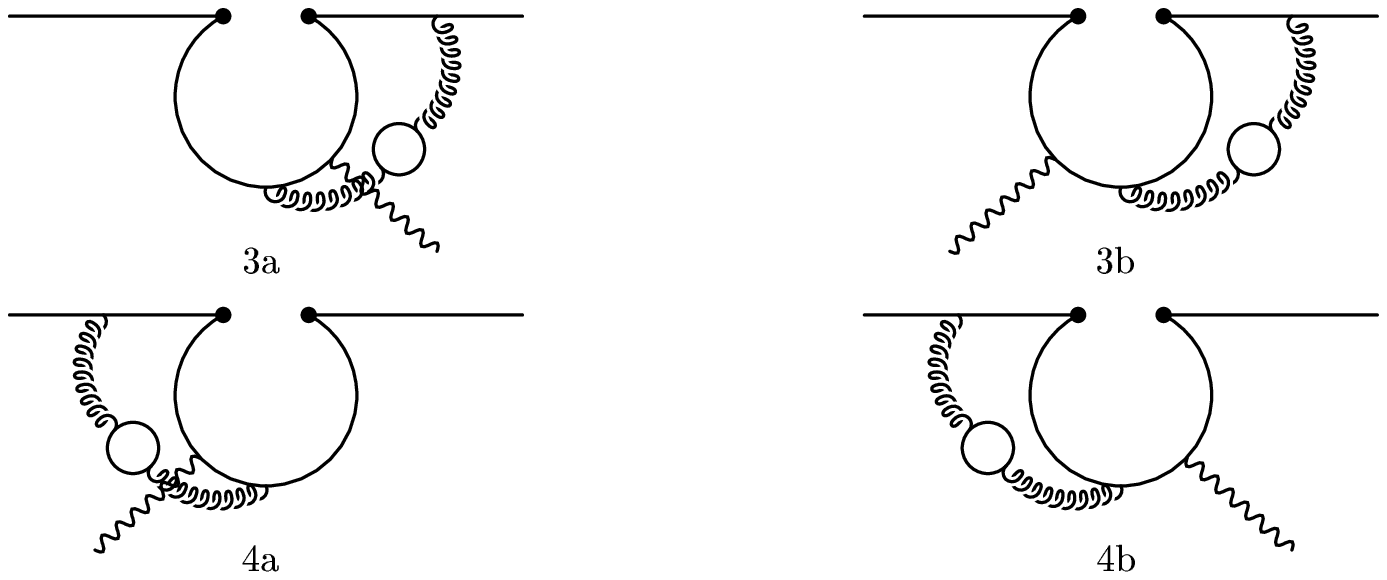}
  \caption{\label{fig:M3M4} Diagrams 3a--b and 4a--b associated with the
    operator $O_2$. The photon is represented by a wavy line and is emitted
    from an up-type quark in all the diagrams. The virtual gluons are
    represented by curly lines. The sum of the first two graphs is denoted
    with $M_{2,{\rm bare}}^{(2)}(3)$, 
    whereas the sum of the second two diagrams is
    called $M_{2,{\rm bare}}^{(2)}(4)$.}
\end{figure}

The three-loop diagrams contributing to $M_2^{(2)}$
can be divided into four non-vanishing classes as
shown in Figs.~\ref{fig:M1M2} 
and~\ref{fig:M3M4}\footnote{In principle there are also diagrams in which 
  the photon
  is emitted from the quark-loop insertion in the gluon propagator.
  However, these contributions vanish due to Furry's theorem.}.
The sum of the diagrams in
each class is gauge invariant. The contributions to the matrix element
$M_2^{(2)}$ 
of the individual classes are denoted by $M_{2,{\rm bare}}^{(2)}(1), \eqnspc
M_{2,{\rm bare}}^{(2)}(2), \eqnspc M_{2,{\rm bare}}^{(2)}(3)$ and $M_{2,{\rm
bare}}^{(2)}(4)$, where, e.g., $M_{2,{\rm bare}}^{(2)}(1)$ is
\begin{equation}
  M_{2,{\rm bare}}^{(2)}(1) = M_{2,{\rm bare}}^{(2)}(1a) + M_{2,{\rm
  bare}}^{(2)}(1b) + M_{2,{\rm bare}}^{(2)}(1c)\,.
\end{equation}

For the practical calculation we essentially follow the techniques
developed in~\cite{Greub:1996tg}. To make the paper self-contained,
we nevertheless present as an example
the calculation of the diagram 2c in some detail. 

The amplitude $M_{2,{\rm bare}}^{(2)}(2c)$
is constructed with the help of the building blocks 
$I_\beta$ and $K_{\beta\beta^\prime}^f$, shown in Fig.~\ref{fig:build1}
in Appendix~\ref{sec:build}. The analytic expression for $I_\beta$
is given in Eq.~(\ref{eq:build1}), while $K_{\beta\beta^\prime}^f$
is given in Eq.~(\ref{eq:build3})
for an arbitrary mass $m_f$ of the quark in the loop.
This mass is retained
in $K_{\beta\beta^\prime}^f$, because it will be used as a regulator
of infrared singularities in the calculation of 
$\langle s \gamma|O_7|b \rangle$. As $\langle s \gamma|O_2|b \rangle$
is free of infrared singularities, we can put in this section $m_f=0$.
Thus the parameter integral in Eq.~(\ref{eq:build3}) can be 
expressed in terms of Euler $\Gamma$ functions.
Furthermore, only the $g_{\beta\beta^\prime}$ term has to be kept as the
other building block $I_\beta$ is transversal.
The diagram 2c can be written as
\bea
M_{2,{\rm bare}}^{(2)}(2c) 
& = & -\frac{2i}{(4\pi)^{\e}} \left(\frac{\a_s}{\pi}\right)^2 e Q_d C_F
T n_f 
\frac{\Gamma^2(\e) \Gamma^2(2-\e)(1-\e)}{\Gamma(4-2\e)} 
\nonumber \\
&& e^{2i\pi\e+3\g_E \e} \m^{6\e} \int \, \frac{{\rm d}^dr}{(2\pi)^d} \, 
\bar{u}(p') \left(r_{\b} \dslash{r}-r^2 \g_{\b}  \right) \nonumber \\
&&  L
\frac{\dslash{p}'+\dslash{r}+m_b}{(p'+r)^2-m_b^2+i \delta}
\dslash{\varepsilon}
\frac{\dslash{p}+\dslash{r}+m_b}{(p+r)^2-m_b^2 + i \delta}\g^{\b} u(p) 
\frac{1}{(r^2+ i\delta)^{1+\e}} \nonumber \\
&& \int_0^1 \, {\rm d}x \, x^{1-\e}(1-x)^{1-\e} \left(r^2-\frac{m_c^2}
{x(1-x)}+i\delta \right)^{-\e}
\,,
\label{eq:mat1}
\eea
where $u(p)$ and $u(p')$ are the Dirac spinors of the $b$
and $s$ quark, respectively, while $\varepsilon$ denotes the 
polarization vector of the photon. 
$C_F$ and $T$ are the eigenvalue of the quadratic Casimir
operator and the index of the fundamental representation of the color
gauge group, respectively, with the numerical values
$C_F \, = \, 4/3$ and $T\,=\,1/2$. The Euler constant $\gamma_E$
appears in Eq.~(\ref{eq:mat1}), because we write the square of the
renormalization scale in the form $\mu^2 \exp(\gamma_E)/(4\pi)$.
The parameter $\delta$ (with $\delta>0$) in the denominators of the various 
propagators symbolizes the ``$\e$-prescription''.

In a next step we denote the four different denominators with
\begin{eqnarray*}
D_1 & = & (p'+r)^2 - m_b^2 + i \, \delta, \\
D_2 & = & (p+r)^2 - m_b^2 + i \, \delta, \\
D_3 & = & r^2 - \frac{m_c^2}{x(1-x)} + i \, \delta , \\
D_4 & = & r^2 + i \, \delta,
\end{eqnarray*}
and introduce a Feynman parametrization as follows:
\bea
\frac{1}{D_1 D_2 D_3^{\e} D_4^{1+\e}} & = & \frac{\Gamma(3+2\e)}{\Gamma(\e)
\Gamma(1+\e)} \int \, 
\frac{{\rm d}u\, {\rm d}v\, {\rm d}y \,w^{\e}y^{\e-1}}{(D_1 u + D_2 v + D_3 y 
+ D_4 w)^{3+2\e}},
\eea
with $w=1-u-v-y$. The integration variables ($u, \, v$ and $y$) run in
the simplex $S$ defined through
$u, \, v, \, y \geq 0$ and $u+v+y \leq 1$. After 
the integration over $r$ one simplifies the remaining integrals with
the help of the substitutions
\begin{equation}
  u \rightarrow (1-u') \left(1-\frac{1-v'}{u'} \right)\,, \, \, v \rightarrow
  \frac{1-u'}{u'}(1-v')\,, \, \, y \rightarrow u'y'\,.
\end{equation}
The integration variable $v'$ varies in the interval $[1-u',\,1]$ whereas the
other three variables $x,\, y'$ and $u'$ all vary in the interval
$[0,\,1]$. We tighten the notation by omitting the primes and arrive at
\begin{eqnarray}
\label{m2cdec}
  M_{2,{\rm bare}}^{(2)}(2c) 
  & = & \frac{1}{8\p^2} \left(\frac{\a_s}{\p}\right)^2 e Q_d C_F T n_f
  \frac{\Gamma(\e) \Gamma^2(2-\e) (1-\e)}{\Gamma(4-2\e)} 
  \nonumber \\
  & & e^{3\g_E \e}
  \m^{6\e}\int_0^1{\rm d}x\int_0^1{\rm d}y\int_0^1{\rm d}u\int_{1-u}^1
  {\rm d}v\,
  x^{1-\e}(1-x)^{1-\e}y^{\e-1} 
  \nonumber \\
  & &  (1-y)^{\e}u^{2\e-1}\bar{u}(p')
  \left(\frac{P_1}{\hat{C}^{1+3\e}}+\frac{P_2}{\hat{C}^{3\e}}
    +\frac{P_3\hat{C}}{\hat{C}^{3\e}}
  \right)u(p)\,, 
\end{eqnarray}
where the Dirac matrices $P_1,\, P_2$ and $P_3$ are polynoms in the Feynman parameters
and the expression $\hat{C}$ is given by
\begin{equation}
  \label{eq:new_prop}
  \hat{C} = m_b^2(1-u)v+\frac{uy}{x(1-x)}m_c^2-i\d\,.
\end{equation}
We should mention at this point that the expression
in Eq.~(\ref{m2cdec}) is infrared finite and is therefore
regularized for $\e>0$.

We use the same approach as in
~\cite{Greub:1996tg,Greub:2000an,Asatryan:2001zw} and 
introduce Mellin-Barnes representations for the denominators $\hat{C}^{1+3\e}$
and $\hat{C}^{3\e}$. In general the Mellin-Barnes representation of an 
expression of the form
$(K^2-M^2)^{-\l}$ (with $\l>0$) reads
\begin{equation}
  \label{eq:mellin}
  \frac{1}{(K^2-M^2)^\l} = \frac{1}{(K^2)^\l}\frac{1}{\Gamma(\l)}\frac{1}{2\p
    i}\int_{\g}{\rm d}s\, 
  \left(-\frac{M^2}{K^2}\right)^s\Gamma(-s)\Gamma(\l+s)\,,
\end{equation}
where the integration path \(\g\) runs parallel to the imaginary axis.
It intersects
the real axis somewhere between $-\l$ and $0$. The Mellin-Barnes
representation for $\hat{C}^\l$, ($\l \in \{3\e, \, 1+3\e\}$)
is implemented by identifying
$K^2$ and $M^2$ as 
\begin{eqnarray}
K^2 & \leftrightarrow & m_b^2(1-u)v\,, \nonumber \\
M^2 & \leftrightarrow & -\frac{uy}{x(1-x)}m_c^2+i\d\,.
  \label{eq:KMdef}
\end{eqnarray}
The integration path \(\g\) has to be chosen such that the parameter integrals
exist for all values of $s \in \g$. This means in our case that 
\(\g\) has to intersect the real $s$-axis between
$-3\e$ and $0$. 
After interchanging the order of
integration, the four Feynman parameter integrals can easily be expressed
in
terms of products of Euler $\Gamma$-functions. What remains to be done is the
integration over \(\g\) in the complex $s$-plane. We close the integration
path in the right half-plane and use the residue theorem to perform this
integral. The residues are located at the following
positions:
\begin{eqnarray}
  s & = & 0, \, 1, \, 2 , \dots \,,\nonumber\\
  s & = & 1-\e, \, 2-\e, \, 3-\e,  \dots \,,\nonumber\\
  s & = & 1-2\e, \, 2-2\e, \, 3-2\e, \dots \,,\nonumber\\
  s & = & 1-3\e, \, 2-3\e, \, 3-3\e, \dots \,,\nonumber\\
  s & = & \tfrac{1}{2}-3\e, \, \tfrac{3}{2}-3\e, \, \tfrac{5}{2}-3\e, 
  \dots\,.
  \label{eq:residue}
\end{eqnarray}
The sum over the residues naturally leads to an expansion 
in the small parameter \(z=m_c^2/m_b^2\) through  
the factor \((m_c^2/m_b^2)^s\) in  
Eq.~(\ref{eq:mellin}) (see also Eq.~(\ref{eq:KMdef})).
This expansion, however, is not a Taylor series because it also
involves logarithms of $z$, which are generated by the expansion in
$\e$.
The final result for $M_{2,{\rm bare}}^{(2)}(2c)$
can thus be written as
\begin{equation}
  \label{eq:end_2c}
  M_{2,{\rm bare}}^{(2)}(2c) = \sum_{k,l} f_{k,l} z^k \ln^l(z),
\end{equation}
where the coefficients $f_{k,l}$ are independent of $z$.
The power $k$ is an (non-negative) integer multiple of 
\(\tfrac{1}{2}\) and $l \in \{0,
1,2,3,4\}$. For a detailed explanation of the range of $l$ we refer
to~\cite{Greub:1996tg}. 

In a similar way all other diagrams can be treated. The final result for the
sum of the three-loop diagrams is given by
\begin{eqnarray}
  M_{2,{\rm bare}}^{(2)} &=& 
  M_{2,{\rm bare}}^{(2)}(1) + M_{2,{\rm bare}}^{(2)}(2) + M_{2,{\rm
  bare}}^{(2)}(3) + M_{2,{\rm bare}}^{(2)}(4)
  \,,
\end{eqnarray}
where the analytical results for the individual terms of the r.h.s.
are listed in Appendix~\ref{sec:threeloopresults}. 
We decided to include corrections up to ${\cal O}(z^3)$ as the 
higher order terms lead to a negligible contribution
for the physical value $z\approx0.1$.

\subsection{Counterterm contributions to $\bm{\langle s \gamma|O_2|b \rangle}$}
In this section we work out the various counterterms
of order $\alpha_s^2 n_f$
which are needed to derive the renormalized result $M_2^{(2)}$.
There are counterterm contributions 
due to the renormalization of the strong
coupling constant
and due to the mixing of $O_2$ into other operators.

We first discuss the counterterms related to the renormalization 
of $\as$.
As the leading term $M_{2,{\rm bare}}^{(0)}$ is zero, only the 
renormalization of $g_s$ in the two-loop result 
$M_{2,{\rm bare}}^{(1)}$ generates a counterterm
which can be written as
\begin{eqnarray}
  M_{2,g_s}^{(2)} &=& 
  2 \delta Z_{g_s}^{(1),n_f} M_{2,{\rm bare}}^{(1)}\,,
  \nn\\
  \delta Z_{g_s}^{(1),n_f} &=& \frac{\a_s}{\p}\frac{n_f T}{6\e}\,.
  \label{eq:m_gs2}
\end{eqnarray}
$M_{2,{\rm bare}}^{(1)}$ 
is the sum of the two-loop diagrams which has to be known
including terms of ${\cal O}(\e)$. 
For this reason we extended the calculation of Ref.~\cite{Greub:1996tg}
to order $\epsilon^1$.

\begin{figure}[t]
  \centering
  \includegraphics[height=2.9cm,angle=0]{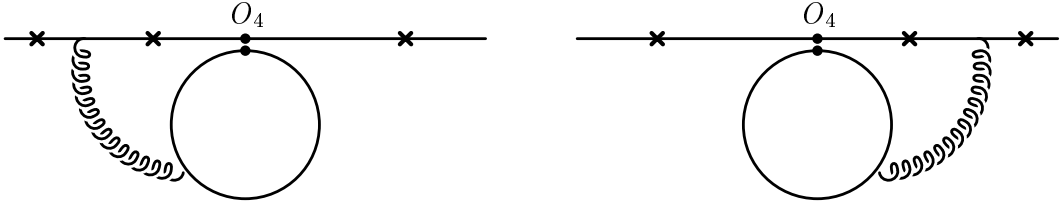}
  \caption{\label{fig:ovier} Counterterm diagrams to $O_2$ involving the
  operator $O_4$. The crosses denote the possible places for photon
  emission. Note that the diagrams where the photon is emitted 
  from the fermion-loop are zero due to Furry's theorem.}
\end{figure}
We now turn to the counterterms induced through the mixing of $O_2$
with other operators. First, we consider the counterterms connected with
the mixing of $O_2$ into
four-fermion operators. At order $\as$ there are non-vanishing mixings
into $O_1$, $O_4$ and into the evanescent operator $P_{11}$, defined
in Appendix A of Ref.~\cite{Chetyrkin:1996vx}. As only $O_4$
has a non-vanishing matrix element 
for $b \to s \gamma$ proportional to $\as n_f$,
the only counterterm of this type is
\begin{eqnarray}
  \label{eq:oviera}
  M_{24,a}^{(2)} &=& \delta Z_{24}^{(1)} M_4^{(1)} \,, \nn \\ 
  \delta Z_{24}^{(1)} &=& \frac{\alpha_s}{\p} \frac{1}{6\e}\,, \nn \\
  M_4^{(1)} &=& \frac{1}{81} \Bigg( -\frac{72}{\e} + 78 + 288 \mub + 36 i \p +
  1159\e - 150\p^2\e \nn \\ && - 312 \ln\left(\frac{m_b}{\m}\right)\e - 576
  \ln^2\left(\frac{m_b}{\m}\right)\e + 258 i \p\e - 144 i \p \mub \e +
  \ordo{\e^2}\Bigg) \nn \\ &&
  \frac{\alpha_s}{4\p} C_F T n_f Q_d \bra s\gamma|O_7|b\ket_\text{tree}\,, 
\end{eqnarray}
where $\delta Z_{24}^{(1)}$ can be found in~\cite{Chetyrkin:1996vx}. 
The Feynman diagrams
contributing to $M_4^{(1)}$, i.e. to the corrections of ${\cal
  O}(\alpha_s n_f)$ to $\bra s \gamma|O_4|b\ket_\text{tree}$,
are depicted in \Fref{fig:ovier}.
They were computed following the strategy outlined in
Section~\ref{sec:threeloop}. 

At order $\as^2 n_f$, there are mixings of $O_2$ into $O_1$, $O_4$
and $P_{11}$ and again  only $O_4$ has a matrix element of ${\cal O}(\as^0)$.
Thus the only counterterm of this type reads
\begin{eqnarray}
  \label{eq:ovierb}
  M_{24,b}^{(2)} &=& \delta Z_{24}^{(2),n_f} M_4^{(0)} \,, \nn \\ 
  \delta Z_{24}^{(2),n_f} &=& \left(\frac{a_s}{\p}\right)^2
  \frac{n_f T}{18\e^2}\,, \nn \\
  M_4^{(0)} &=& \Big(1-2\mub \e + \frac{\p^2\e^2}{12} +
  2\ln^2\left(\frac{m_b}{\m}\right)\e^2 + \ordo{\e^3}\Big) C_F Q_d \bra
  s\gamma|O_7|b\ket_\text{tree}\,.
\end{eqnarray}

In a second step we consider the counterterms connected with
the mixing of $O_2$ into the dipole operators $O_7$ and $O_8$. 
One can easily see that only one counterterm of this type
generates a contribution of ${\cal O}(\as^2 n_f)$: $O_2$ mixes
at three-loop order 
into $O_7$; in turn, from $O_7$ the tree-level matrix element
for $b \to s \gamma$ is taken. The resulting counterterm therefore reads
~\cite{Chetyrkin:1996vx,MMpriv}
\begin{eqnarray}
  \label{eq:o27}
  M_{27}^{(2)} 
  &=& \delta Z_{27}^{(2),n_f} \bra s \gamma|O_7|b\ket_\text{tree}\,,
  \nonumber\\ 
  \delta
  Z_{27}^{(2),n_f} &=& \left(\frac{\alpha_s}{\p}\right)^2 C_F T n_f
  \left[\frac{1}{\e^2} \left(\frac{Q_u}{24}-\frac{Q_d}{81} \right) -
  \frac{1}{\e}\left(\frac{Q_u}{144}+\frac{2Q_d}{243} \right) \right]\,,
\end{eqnarray}
where $Q_u=2/3$ and $Q_d=-1/3$ are the charge factors of up- and 
down-type quarks, respectively.

\subsection{Renormalized result for $\bm{\langle s \gamma|O_2|b \rangle}$}
Combining the three-loop result $M_{2,\rm bare}^{(2)}$, calculated
in Subsection~A, with the various counterterm contributions discussed
in Subsection~B (see Eqs.~(\ref{eq:m_gs2}),
(\ref{eq:oviera}), (\ref{eq:ovierb}),
 and (\ref{eq:o27})), we get an ultraviolet finite result. 
As mentioned earlier, the result is also free of 
infrared singularities.
Inserting the numerical 
values for the color factors
($C_F=4/3$, $T=1/2$) and the electric charge factors ($Q_u=2/3$, $Q_d=-1/3$),
we get the following renormalized result
\begin{eqnarray}
  M_2^{(2)} &=& M_{2,{\rm bare}}^{(2)} 
  + M_{2,g_s}^{(2)} + M_{24,a}^{(2)} +  M_{24,b}^{(2)} +
M_{27}^{(2)}
  \nonumber\\
  &=&
  \left(\frac{\alpha_s}{4\p}\right)^2 n_f \, \bra s\gamma|O_7|b\ket_\text{tree}
  \left(t_2^{(2)} \mubb +l_2^{(2)} \mub + r_2^{(2)} \right) \,,
  \label{eq:M2}
\end{eqnarray}
with
\begin{eqnarray}
  \label{eq:l2}
  t_2^{(2)} &=& \frac{800}{243}\,, \\ 
  \Re\left(l_2^{(2)}\right) &=&\frac{16}{243}
  \Big( -145 + \left(288-30\p^2 -216 \zeta(3) + 216 L - 54\p^2 L + 18
  L^2 \right. \nn \\ &&\left. + 6 L^3 \right) z + 24 \p^2 z^{3/2} +
  6 \left(18 + 2 \p^2 +12 L -6\p^2 L + L^3\right) z^2 \nn \\ && -\left(9 +
  14\p^2 -182L + 126L^2\right)z^3 \Big) + {\cal O}(z^4)\,, \\
  \Im\left(l_2^{(2)}\right) &=& \frac{16\p}{243} \Big(-22 + \left(180 -12\p^2
  + 36L + 36L^2 \right) z \nn \\ && -\left(12\p^2 - 36 L^2\right) z^2 +
  \left( 112 - 48 L\right) z^3 \Big) +{\cal O}(z^4)\,,
\end{eqnarray}
\begin{eqnarray}
  \label{eq:r2}
  \Re\left(r_2^{(2)}\right) &=& \frac{67454}{6561} - \frac{124\p^2}{729} -
  \frac{4}{1215} \left(11280 - 1520\p^2 - 171\p^4 - 5760 \zeta(3) \right. \nn
  \\ && \left. + 6840L - 1440\p^2L - 2520\zeta(3)L + 120L^2 + 100 L^3 - 30
  L^4\right) z \nn \\ && - \frac{64\p^2}{243} \left( 43 - 12 \ln(2) -
  3L\right) z^{3/2} - \frac{2}{1215} \left(11475 - 380\p^2 + 96\p^4
  \right. \nn \\ && \left. +7200 \zeta(3) - 1110L - 1560\p^2L + 1440
  \zeta(3)L +990L^2 +260L^3 \right. \nn \\ &&\left.  -60L^4\right)z^2 +
  \frac{2240\p^2}{243} z^{5/2} - \frac{2}{2187} \left(62471 -2424\p^2
  - 33264\zeta(3) \right. \nn \\ && \left. - 19494L - 504\p^2L -5184L^2
  +2160L^3\right) z^3 +{\cal O}(z^{7/2})\,, \\
  \Im\left(r_2^{(2)}\right) &=& \frac{4\p}{729} \Big( 495 - 12\left(375
  -19\p^2 + 36\zeta(3) +84L + 48L^2 -6L^3\right) z \nn \\ &&+6
  \left(207 +38\p^2 -72 \zeta(3) -126L -78L^2 +12L^3\right) z^2 \nn \\
  &&+ 8\left(67 -12\p^2 -48L\right)z^3 \Big) +{\cal O}(z^4)\,,
\end{eqnarray}
where $L=\ln z$.
We note that in the derivation of this ${\cal O}(\as^2 n_f)$ result, 
there was no need to renormalize the parameter $m_b$ in the corresponding
${\cal O}(\as^1)$ expression. Therefore, the symbol 
$\bra s\gamma|O_7|b\ket_\text{tree}$
can be interpreted to be
(in $M_2^{(1)}$ and $M_2^{(2)}$)    
\begin{eqnarray}
  \bra s\gamma|O_7|b\ket_\text{tree} &=& 
   m_b\frac{e}{8\pi^2}
  \bar{u}(p^\prime) \varepsilon\hspace{-0.45em}/ q\hspace{-0.45em}/ u(p) 
  \label{eq:MEO7} \,,
\end{eqnarray}
where $m_b$ denotes the pole mass of the $b$ quark. Concerning this
point, the reader is also referred to Section~\ref{sec:o7calc}.

We now turn to the renormalized matrix element $M_1^{(2)}$, associated with
the operator $O_1$. $O_1$, defined in Eq.~(\ref{opbasis}), can be written as
\beq
O_1 = \frac{1}{2} \, \widetilde{O}_1 - \frac{1}{6} \, O_2 \,,
\eeq
with
\beq
 \widetilde{O}_1 = 
 (\bar{s}_L^{\alpha} \gamma_\mu  c_L^{\beta}) \,
 (\bar{c}_L^{\beta} \gamma^\mu  b_L^{\alpha}) \, , 
\eeq
where $\alpha$ and $\beta$ are color indices. It is easy to
see that $\widetilde{O}_1$ has a vanishing matrix element for
$b \to s \gamma$. Therefore, one obtains 
\beq
M_1^{(2)} = - \frac{1}{6} \, M_2^{(2)} \, .
\eeq

\section{Real and virtual corrections to $\bm{\langle s \gamma|O_7|b \rangle}$}
\label{sec:o7calc}

\begin{figure}[t]
  \begin{center}
    \leavevmode
    \includegraphics[height=3.2cm,angle=0]{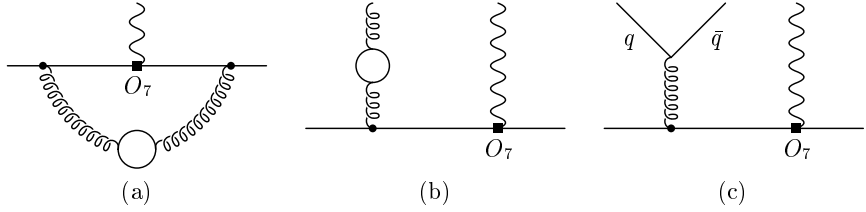}
  \end{center}
  \caption{\label{fig:osieben} Virtual (a), gluon-bremsstrahlung (b) and
    quark-pair radiation (c) graphs to the operator $O_7$. In (b) and (c), the
    diagrams where the gluon is emitted from the $s$-quark are not
    shown.}
\end{figure}
In this section we describe in some detail the steps needed for the
calculation of the
${\cal{O}}(\a_s^2 n_f)$ corrections to the matrix element
 $\langle s \gamma|O_7|b \rangle$.
Due to the presence of infrared singularities, 
the practical calculation
proceeds in a slightly different way than for $O_2$.
As these singularities only get canceled at the level of the decay width
when combining the virtual corrections shown in  
Fig.~\ref{fig:osieben}(a) with the gluon bremsstrahlung 
(Fig.~\ref{fig:osieben}(b)) and the
quark-pair emission process (Fig.~\ref{fig:osieben}(c)),
we first derive expressions for the ${\cal O}(\as^2 n_f)$
corrections to these three contributions to the decay width.
The corresponding expressions necessary to evaluate 
${\rm BR}(B \to X_s \gamma)_{E_\gamma\ge E_{\rm cut}}$
are discussed in Appendix~\ref{sec:cutoff}.

To fix the notation, we write the contribution 
from $O_7$ to the decay width $\Gamma(b\to X_s\gamma)$ as
\beq
\label{gammadecomp}
  \Gamma_{77} = \Gamma_{77}^0 \left[ 1+
\hat{\Gamma}_{77}^{(1)} + \hat{\Gamma}_{77}^{(2),n_f} 
\right]\,,\qquad
  \Gamma_{77}^0 = \frac{m_b^5\alpha_{{\rm em}}}{32\p^4} 
  |G_F\lambda_t C_7^{\text{eff}}|^2
  \,.
\eeq
The ${\cal O}(\as)$ correction, $\hat{\Gamma}_{77}^{(1)}$,
can be extracted from Ref.~\cite{Greub:1996tg}, reading
\beq
\hat{\Gamma}_{77}^{(1)} = \frac{\as}{4\pi}
\left( -\frac{32}{9} - \frac{16\pi^2}{9} 
 + \frac{64}{3} \mub \right) \, .
\eeq
We further split $\hat{\Gamma}_{77}^{(2),n_f}$ in Eq.~(\ref{gammadecomp})
as
\beq
\hat{\Gamma}_{77}^{(2),n_f} =
\hat{\Gamma}_{77}^{(2),(a)} +
\hat{\Gamma}_{77}^{(2),(b)} +
\hat{\Gamma}_{77}^{(2),(c)} \, ,
\eeq
with obvious notation (\Fref{fig:osieben}).

For the calculation of the three parts contributing to
$\hat{\Gamma}_{77}^{(2),n_f}$ we could in principle 
put $m_f=m_s=0$ at the beginning of the
calculation and use dimensional regularization for both infrared
and ultraviolet singularities. We found it easier, however,
to use the strange quark mass, $m_s$, and the mass of the quark in the
fermion bubble,
$m_f$, as infrared regulators. 
For formulating the results, it is convenient to introduce the
dimensionless quantities 
\beq
r = \frac{m_s^2}{m_b^2} \,,\qquad
f = \frac{m_f^2}{m_b^2} \,.  
\eeq
We now turn to the
calculations of 
$\hat{\Gamma}_{77}^{(2),(c)}$, 
$\hat{\Gamma}_{77}^{(2),(b)}$ and 
$\hat{\Gamma}_{77}^{(2),(a)}$ (in this order). 

Inspecting
the explicit expressions for the quark-pair radiation process
(cf. Fig.~\ref{fig:osieben}(c)), one finds that it
can be worked out in our ``massive''
regularization scheme in $d=4$ dimensions. 
Furthermore, one realizes that one can also put $m_s=0$, provided $m_f$ is
kept at a (small) fixed value. As a consequence, the quark-pair
radiation process is completely regularized by the mass $m_f$.
The evaluation of this process
is quite standard: in a first step the subprocess
$b\to s\gamma g^\star$ is considered where $g^\star$ represents a virtual
gluon. Subsequently the other subprocess, describing the decay 
of $g^\star$ into 
two fermions, is added. It is straightforward to perform the 
occurring phase space integrations where only the one over the gluon
virtuality is non-trivial. 
However, in the limit $m_f\to0$ also this one can be performed
analytically. One arrives at the following result
for the quark-pair emission process:
\begin{eqnarray}
  \label{gamma77c}
  \hat{\Gamma}_{77}^{(2),(c)} &=& \left( \frac{\as}{4\pi}\right)^2  
  \frac{n_f}{243}
  \left[ - 12662 + 24\p^2 +  2592\zeta(3) +
    (144\p^2 - 5916)\ln(f) \right. \nn \\ && \left.-900 \ln^2(f) -72\ln^3(f)
     \right]\,.
\end{eqnarray}
Due to the Kinoshita-Lee-Nauenberg theorem, 
it follows that the sum of the virtual
and the gluon bremsstrahlung corrections
also must be finite for $d \to 4$ and $m_s \to 0$ for fixed $m_f$.

We now turn to the gluon bremsstrahlung process. The diagram
in Fig.~\ref{fig:osieben}(b) (combined with the one where the gluon is emitted
from the $s$-quark) can be written as
\beq
M_{7,{\rm bare}}^{(2),(b)} = \frac{\delta Z_3^{(1),n_f}}{2} 
\, M_{7}^{(1),(b)}
\,,
\eeq
where $M_{7}^{(1),(b)}$ denotes the lowest order matrix element
for $b \to s \gamma g$ and $\delta Z_3^{(1),n_f}$ reads
\beq
  \delta Z_3^{(1),n_f} = -\frac{\alpha_s}{\pi} \frac{n_f T}{36} \left(
  \frac{12}{\epsilon} -24\ln\left(\frac{m_f}{\mu}\right) + \pi^2\epsilon
  +24\ln^2\left(\frac{m_f}{\mu}\right)\epsilon + {\cal O}(\e^2) \right)\, .
\eeq 
Note that the $1/\e$ pole is of ultraviolet origin;
the infrared singularity is regulated by $m_f$ in this expression. 
In addition, there is a counterterm contribution due to the 
$\overline{\rm MS}$
renormalization of the strong coupling constant of the form
\beq
M_{7,{\rm ct}}^{(2),(b)} = \delta Z_{g_s}^{(1),n_f} \, M_{7}^{(1),(b)}
\,,
\eeq
with
\beq
  \delta Z_{g_s}^{(1),n_f} = \frac{\alpha_s}{\pi}\frac{n_f T}{6\epsilon}\,.
\eeq
Combining $M_{7,{\rm bare}}^{(2),(b)}$ with $M_{7,{\rm ct}}^{(2),(b)}$,
one obtains the renormalized matrix element $M_{7}^{(2),(b)}$
\beq
M_{7}^{(2),(b)} = 
\left(\delta Z_{g_s}^{(1),n_f} + 
\frac{\delta Z_3^{(1),n_f}}{2}\right) 
M_{7}^{(1),(b)} \,,
\eeq
from which 
the ${\cal O}(\as^2 n_f)$ contribution to the decay width is
obtained in a straightforward way. One gets
\begin{eqnarray}
\label{gamma77b}
 \hat{\Gamma}_{77}^{(2),(b)} &=& 2 \left(\delta Z_{g_s}^{(1),n_f} + 
 \frac{\delta Z_3^{(1),n_f}}{2}\right) \hat{\Gamma}_{77}^{(1),(b)} \nonumber \\
  &=& \left(\frac{\a_s}{4\p}\right)^2\frac{C_F T n_f}{18}\Bigg[
    \frac{48}{\epsilon}\left(2\ln(f)+4\ln\left(\frac{m_b}{\mu}\right)+
    \ln(f)\ln(r)\right. \nonumber \\
  && \left. +2 \ln\left(\frac{m_b}{\mu}\right)\ln(r)\right) -8\pi^2
  +416\ln(f) -32\pi^2\ln(f) \nonumber \\
  && -48\ln^2(f) +832\ln\left(\frac{m_b}{\mu}\right) - 64\pi^2
  \ln\left(\frac{m_b}{\mu}\right) -960\ln^2\left(\frac{m_b}{\mu}\right)
  \nonumber \\ 
  && -576\ln(f) \ln\left(\frac{m_b}{\mu}\right) -4\ln(r)\Bigg( \pi^2 -18 \ln(f)
  + 6\ln^2(f)  \nonumber \\
  && -36\ln\left(\frac{m_b}{\mu}\right)+120\ln^2\left(\frac{m_b}{\mu}\right)
  +72\ln(f)\ln\left(\frac{m_b}{\mu}\right)\Bigg) \nonumber \\
  && -24\ln^2(r)\left(\ln(f)+2\ln\left(\frac{m_b}{\mu}\right)\right)\Bigg]\,,
\end{eqnarray}
where $\hat{\Gamma}_{77}^{(1),(b)}$ is the corresponding 
(normalized) decay width for
$b \to s \gamma g$ in the ${\cal O}(\as)$ approximation. As in our
regularization scheme the sum
$\delta Z_{g_s}^{(1),n_f} + \delta Z_3^{(1),n_f}/2$ is finite (in $\e$), 
$\hat{\Gamma}_{77}^{(1),(b)}$ is only needed up to terms of order $\e^0$,
which simplified the calculation.

We now turn to the evaluation of the virtual
corrections shown
in Fig.~\ref{fig:osieben}(a)
and also discuss the various counterterm contributions.
For the diagram shown in this figure, we obtain
\begin{eqnarray}
\label{m7virtvertex}
  M_{7,{\rm bare}}^{(2),(a)} &=&
  \frac{1}{81}\Bigg[\frac{54}{\epsilon^2}\left(2\ln(r)-1\right)
  +\frac{18}{\epsilon} \bigg(2+ 12\ln(r) -6\ln(r)\ln(f)\nonumber\\
  && -24\ln(r)\mub-3\ln^2(r) +6\ln(f)+ 12\mub\bigg) \nonumber \\
  && +1718 + 123\pi^2 +840\ln(f) +36\pi^2\ln(f) +90\ln^2(f) \nonumber \\
  && +18\ln^3(f) -144\mub -432\mubb \nonumber \\
  &&-432\ln(f)\mub +18\ln(r)\bigg( 24 +\pi^2 -12\ln(f) +3\ln^2(f) \nonumber \\
  &&-48\mub + 48\mubb +24\ln(f)\mub \bigg) \nonumber \\
  && -54\ln^2(r)\left(2-\ln(f)-4\mub\right) +18\ln^3(r) \Bigg] \nonumber \\
  && \left(\frac{\alpha_s}{4\pi}\right)^2 C_F T n_f \bra s \gamma | O_7 | b
  \ket_{\rm tree}\,.
\end{eqnarray}
We stress that this expression is derived in such a way that
$m_s$ is understood to be sent to zero
prior to $m_f$. This procedure is justified by the fact that
for fixed $m_f$ the sum of the virtual- and gluon bremsstrahlung
contributions must be finite in the limit $m_s \to 0$, as discussed
above.

The counterterm contribution $ M_{7,{\rm ct}}^{(2),(a)}$ at
${\cal O}(\as^2 n_f)$ has various sources. There is a contribution
$ M_{7,{\rm ct}_1}^{(2),(a)}$
due to the renormalization of $g_s$ in the ${\cal O}(\a_s)$ 
vertex diagram (i.e. like the one in Fig.~\ref{fig:osieben}(a),
but without the fermion bubble), yielding
\begin{eqnarray}
\label{m7virtcountone}
  M_{7,{\rm ct}_1}^{(2),(a)} 
  &=& \frac{1}{9}\Bigg[-\frac{12}{\epsilon^2}\ln(r) -
  \frac{6}{\epsilon} \ln(r)\bigg(4 -\ln(r) -4\mub\bigg) \nonumber \\
  && +12 -\ln(r)\left(48 +\pi^2 -48\mub +24\mubb\right) \nonumber \\
  && +12\ln^2(r)\left(1-\mub\right) -2\ln^3(r)\Bigg] \nonumber \\
  && \left(\frac{\alpha_s}{4\pi}\right)^2 C_F T n_f \bra s \gamma | O_7 | b
  \ket_{\rm tree}\,.
\end{eqnarray}
Then, there is a counterterm contribution $ M_{7,{\rm ct}_2}^{(2),(a)}$
of the form
\beq
\label{m7virtcounttwo}
  M_{7,{\rm ct}_2}^{(2),(a)} = \left( 
\frac{\delta Z_{2,b}^{(2),n_f}}{2} +
\frac{\delta Z_{2,s}^{(2),n_f}}{2} +
\delta Z_{77}^{(2),n_f} +
\delta Z_{m_b}^{{\rm on},(2),n_f} \right) \, 
\bra s \gamma | O_7 | b \ket_{\rm tree}\,.
\eeq
Here, 
$\delta Z_{2,b}^{(2),n_f}$ and
$\delta Z_{2,s}^{(2),n_f}$ are the ${\cal O}(\as^2 n_f)$ pieces
of the on-shell wave function renormalization constants for the
$b$ and $s$ quark, respectively, while the operator renormalization
factor $\delta Z_{77}^{(2),n_f}$ refers to the $\overline{\rm MS}$ scheme.
Note that the presence of the {\it on-shell} renormalization
factor $\delta Z_{m_b}^{{\rm on},(2),n_f}$ in Eq.~(\ref{m7virtcounttwo})
implies that in the lower order contributions the symbol
$\bra s \gamma | O_7 | b \ket_{\rm tree}$ is understood to be
the tree-level matrix element of $O_7$ in which the running $b$-quark
mass is replaced by the corresponding pole mass. The explicit form
of the various $\delta Z$ factors occurring in Eq.~(\ref{m7virtcounttwo})
can be seen in Appendix~\ref{sec:zfactors}.

After combining Eqs.~(\ref{m7virtvertex}), (\ref{m7virtcountone}) and
(\ref{m7virtcounttwo}) into the renormalized matrix element, the
calculation of $\hat{\Gamma}_{77}^{(2),(a)}$
is straightforward. We obtain
\begin{eqnarray}
\label{gamma77a}
\hat{\Gamma}_{77}^{(2),(a)} 
  &=& \left( \frac{\as}{4\pi}\right)^2
\frac{C_F T n_f}{81}\Bigg[
    \frac{-216}{\epsilon}\left(2\ln(f)+4\ln\left(\frac{m_b}{\mu}\right)+
    \ln(f)\ln(r)\right. \nonumber \\
  && \left. +2 \ln\left(\frac{m_b}{\mu}\right)\ln(r)\right) +7495 + 624\pi^2 +
  1086\ln(f) + 72\pi^2\ln(f) \nonumber \\
  && +666\ln^2(f) + 36\ln^3(f) -6336\mub +6048\mubb \nonumber \\
  && + 2592\ln(f)\mub +18\ln(r)\Bigg( \pi^2 -18 \ln(f)
  + 6\ln^2(f)  \nonumber \\
  && -36\mub+120\mubb +72\ln(f)\mub \Bigg) \nonumber \\
  && +108\ln^2(r)\left(\ln(f)+2\ln\left(\frac{m_b}{\mu}\right)\right)\Bigg]\,.
\end{eqnarray}
We now combine virtual and gluon bremsstrahlung corrections given
in Eqs.~(\ref{gamma77a}) and (\ref{gamma77b}), respectively. We obtain
(after putting $T=1/2$ and $C_F=4/3$)
\begin{eqnarray}
  \label{gamma77ab}
  \hat{\Gamma}_{77}^{(2),(a)+(b)} &=& 
  \left(\frac{\a_s}{4\p}\right)^2
  \frac{n_f}{243}
  \Big[ 14990 + 1176\p^2 + (5916-144\p^2)\ln(f) +
  900 \ln^2(f) 
  \nn \\ && + 72\ln^3(f) -576(9+\p^2) \mub + 3456 \mubb\Big]\,, 
\end{eqnarray}
where the $1/\e$ poles and the mass singularities associated with $m_s$
are canceled.

When combining this result with 
the quark-pair emission process in Eq.~(\ref{gamma77c}), we obtain
the final result
\begin{eqnarray}
  \label{eq:o7_final}
  \hat{\Gamma}_{77}^{(2),n_f} 
  &=& \left(\frac{\a_s}{4\p}\right)^2 n_f
  \left( 
    2 t_7^{(2)} \mubb + 2 l_7^{(2)} \mub + 2r_7^{(2)} 
  \right)
  \,,
\end{eqnarray}
with
\begin{eqnarray}
  t_7^{(2)} &=& \frac{64}{9}\,, \nn \\
  l_7^{(2)} &=& -\frac{32}{27}\left(9 +\p^2\right)\,,\nn\\
  r_7^{(2)} &=& \frac{4}{81}\left(97 +50\p^2 +108\zeta(3)\right) \,.
\end{eqnarray}
The cancelation of the $\ln(f)$ terms 
is a strong check for the correctness of the
individual pieces of the calculation.

For later convenience we formally introduce an amplitude $M_7$ in such
a way that its square reproduces the result of Eq.~(\ref{eq:o7_final}).
Adopting the notation of Eq.~(\ref{eq:M2def}) one gets
\begin{eqnarray}
  M_7^{(2)}
  &=&
  \left(\frac{\alpha_s}{4\p}\right)^2 n_f \bra s\gamma|O_7|b\ket_\text{tree}
  \left( t_7^{(2)} \mubb +l_7^{(2)} \mub +r_7^{(2)} \right)
  \,.
  \label{eq:M7}
\end{eqnarray}

\section{Virtual corrections to $\bm{\langle s \gamma|O_8|b \rangle}$}
\label{sec:o8calc}
\begin{figure}[t]
  \includegraphics[height=3cm]{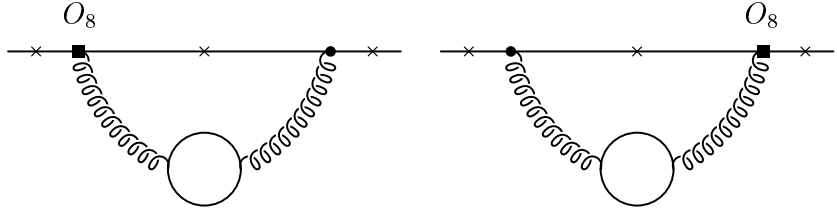}
  \caption{\label{fig:O8virt} Graphs associated with virtual corrections to
    the operator $O_8$. The crosses denote the possible places where the
    photon can be emitted.}
\end{figure}
We first discuss the two-loop diagrams depicted in
\Fref{fig:O8virt}, which contain the building block $K_{\b\b'}^f$ (see
\Eref{eq:build3}). As these diagrams are free of infrared singularities,
we put the masses $m_f$ of the quarks in the fermion loop as well as
the strange quark mass $m_s$
to zero from the beginning.
The calculation can be performed along the same lines as described in
Section~\ref{sec:threeloop}. However, due to the absence of $m_c$, the actual
evaluation of the diagrams turns out to be much simpler.
The result can be cast into the form
\begin{eqnarray}
  \label{eq:o8virt}
  M_{8,{\rm bare}}^{(2)} &=&
  \left(\frac{\a_s}{4\p}\right)^2
  C_F T n_f Q_d \bra s \g | O_7 | b\ket_{\text{tree}}
  \frac{4}{27}\bigg[\left(\frac{18}{\e^2}+\frac{1}{\e}\left(120 - 6\pi^2 +
      18i\pi\right)\right)\left(\frac{m_b}{\mu}\right)^{-4\e} 
  \nn \\ &&
  + 530 - 
  28\pi^2 -180 \zeta(3) + 93i\p\bigg]
  \,.
\end{eqnarray}
The counterterm contribution of ${\cal O}(\as^2 n_f)$,
denoted by $M_{8,{\rm ct}}^{(2)}$, stems 
from the renormalization of $g_s$ and from the
mixing of $O_8$ into the operator $O_7$. We obtain
\beq
\label{m8ct}
    M_{8,{\rm ct}}^{(2)} = \delta Z_{87}^{(2),n_f} 
    \bra s\g|O_7| b\ket_{\text{tree}} + 2\delta Z_{g_s}^{(1),n_f} 
    \, M_8^{(1)} \,, 
\eeq
with
\bea
    \delta Z_{87}^{(2),n_f} &=& \left(\frac{\alpha_s}{\p}\right)^2 C_F T n_f
    \frac{Q_d}{36\e} \left(\frac{6}{\e}-7\right)\, , \nn \\
    \delta Z_{g_s}^{(1),n_f} &=& \frac{\a_s}{\p}\frac{n_f T}{6\e}\, , \nn \\
  M_8^{(1)} &=& -\frac{\alpha_s}{4\pi} \frac{1}{3} Q_d C_F \bra
  s\gamma|O_7|b\ket_{\rm tree}\Bigg[ \frac{12}{\epsilon} +33
  -2\pi^2 - 24\ln\left(\frac{m_b}{\mu}\right)\nonumber \\
  && +6i\pi +\epsilon\left(72-4\pi^2- 36\zeta(3) -66
  \ln\left(\frac{m_b}{\mu}\right) +4\pi^2 \ln\left(\frac{m_b}{\mu}\right)
  \right. \nonumber \\
  && \left. +24\ln^2\left(\frac{m_b}{\mu}\right) +12 i\pi - 12 i\pi
  \ln\left(\frac{m_b}{\mu}\right)\right) + {\cal O}(\e^2) \Bigg] \,.
\eea
$\delta Z_{87}^{(2),n_f}$ 
is obtained from~\cite{Chetyrkin:1996vx,MMpriv}.
The sum of $M_{8,{\rm bare}}^{(2)}$ and $M_{8,{\rm ct}}^{(2)}$ 
 leads to the renormalized result (using $T=1/2$, $C_F=4/3$ and $Q_d=-1/3$)
\begin{equation}
  \label{eq:M8}
  M_8^{(2)}=\left(\frac{\alpha_s}{4\p}\right)^2 n_f 
  \bra s\gamma|O_7|b\ket_{\text{tree}} 
  \left[t_8^{(2)} \mubb + l_8^{(2)} \mub +r_8^{(2)} \right]
  \,,
\end{equation}
with
\begin{eqnarray}
  \label{eq:t8u8v8}
  t_8^{(2)} &=& -\frac{64}{27}\,, \nn \\
  l_8^{(2)} &=& \frac{16}{81}\left(47 - 2\pi^2 + 6i\pi\right)\,, 
  \nn\\ 
  r_8^{(2)} &=& \frac{8}{243}
  \left( -314 + 16\pi^2 + 72\zeta(3) - 57i\pi\right)\,.
\end{eqnarray}

\section{\label{sec:num}Numerical impact of the 
$\bm{{\cal O}(\alpha_s^2 n_f)}$ corrections}
It is well-known that the inclusive decay rate for
$B \to X_s \gamma$ is given by the corresponding $b$-quark 
decay rate $\Gamma(b \to X_s \gamma)$, up to power corrections
of the form $(\Lambda_{{\rm QCD}}/m_b)^2$ \cite{powermb} and
$(\Lambda_{{\rm QCD}}/m_c)^2$ \cite{powermc}
which numerically are well below
$10\%$. 

As our new results are only a part of the complete NNLL
contributions, we do not present a new prediction of the
branching ratio in this paper. Instead, we only illustrate
how the ${\cal O}(\alpha_s^2 n_f)$ corrections to the
matrix elements of the operators $O_1$, $O_2$, $O_7$ and $O_8$
modify the NLL branching ratio for a given set of input parameters.
For this purpose, we neglect power corrections and also
electroweak terms.

In a NLL calculation
the inclusive quark-level transition
$b \to X_s \gamma$ involves the subprocesses
$b \to s \gamma$ (including virtual corrections) and 
$b \to s \gamma g$, i.e., the gluon bremsstrahlung process.
We write the amplitude for the first subprocess similar as in 
Ref.~\cite{BG98}:  
\beq
\label{amplitude}
{\cal A}^{{\rm NLL}}(b \to s \gamma) = - 
 \frac{4 G_F}{\sqrt{2}} \, \lamt \,D^{{\rm NLL}} \langle s \gamma |O_7|b
 \rangle_{\rm tree} \,,
\eeq
where the reduced amplitude $D^{{\rm NLL}}$ reads
\beq
 D^{{\rm NLL}} =  C_7^{{\rm eff}}(\mu)
  + \frac{\as(\mu)}{4 \pi} \, V^{(1)}(\mu) \, . 
\label{drewrite}
\eeq
The symbol $V^{(1)}(\mu)$, defined as 
\beq
  V^{(1)}(\mu) = 
   \sum_{i=1}^8 C_i^{{\rm eff}}(\mu) 
   \left[\left(r^{(1)}_i-\frac{16}{3} \, \delta_{i7}\right)+
         (l^{(1)}_i+ 8 \, \delta_{i7}) \,
             \mub \right] \, ,
\label{vdefine}
\eeq
incorporates the NLL corrections, $r_i^{(1)}$ and 
$l_i^{(1)}$, to the matrix elements.
In Eq.~(\ref{drewrite}), the first term on the r.h.s. is 
understood to be the Wilson coefficient $C_7^{{\rm eff}}(\mu)$
at NLL order, while the Wilson coefficients appearing in 
$V^{(1)}(\mu)$ are understood to be taken at LL order.
As in Ref.~\cite{BG98}, we convert the running mass 
factor $\overline{m}_b(\mu)$, which appears in the definition
of the operator $O_7$ in Eq.~(\ref{opbasis}), into the pole
mass $m_b$. 
This conversion is absorbed into the function $V^{(1)}(\mu)$ and consequently
the symbol 
$ \langle s \gamma |O_7|b \rangle_{\rm tree}$ 
in Eq.~(\ref{amplitude})  
is the tree-level
matrix element of the operator
$O_7$, where the running mass factor $\overline{m}_b(\mu)$ is
understood to be replaced by the pole mass $m_b$.   
The NLL virtual correction functions 
$r_i^{(1)}$ and $l_i^{(1)}$  
in~(\ref{vdefine}), taken from Ref.~\cite{Greub:1996tg},
are repeated for completeness in
Appendix~\ref{sec:rili}. Note, that the quantity $r_7^{(1)}$
not only contains virtual corrections to the matrix element of $O_7$,
which would be infrared singular. $r_7^{(1)}$ is constructed in such a way,
that the $(O_7,O_7)$ interference term generates the sum of virtual
and bremsstrahlung corrections when formally calculating the branching
ratio from ${\cal A}^{{\rm NLL}}(b \to s \gamma)$. 
For the details of this construction, we refer to Ref.~\cite{Greub:1996tg}.
Numerically, the square of this amplitude encodes the bulk of the
decay width. The additional bremsstrahlung corrections, which are 
infrared finite for $E_{\rm{gluon}} \to 0$, 
are relatively small. Therefore, when considering terms 
of order ${\cal O}(\as^2 n_f)$, we omit 
purely finite bremsstrahlung contributions.  

When improving the amplitude for the subprocess $b \to s \gamma$
by including the terms of ${\cal O}(\alpha_s^2 n_f)$,
the result can be written as
\beq
\label{amplitudeimp}
{\cal A}(b \to s \gamma) = - 
 \frac{4 G_F}{\sqrt{2}} \, \lamt \, D
 \langle s \gamma |O_7|b \rangle_{\rm tree} \,,
\eeq
where the reduced amplitude $D$ is 
\beq
 D =  C_7^{{\rm eff}}(\mu)
  + \frac{\as(\mu)}{4 \pi} \, V^{(1)}(\mu) 
  + \left( \frac{\as(\mu)}{4 \pi} \right)^2 \, n_f  \, V^{(2)}(\mu) \, . 
\label{drewriteimp}
\eeq
$V^{(2)}(\mu)$, defined as 
\beq
  V^{(2)}(\mu) = 
   \sum_{i=1}^8 C_i^{{\rm eff}}(\mu) 
   \left[r^{(2)}_i+
         l^{(2)}_i \, \mub  + 
         t^{(2)}_i \, \mubb  \right] \, ,
\label{vdefineimp}
\eeq
incorporates the ${\cal O}(\as^2 n_f)$ corrections to
the matrix elements calculated in the previous sections of this
paper.  The explicit $C_7^{{\rm eff}}(\mu)$ term in Eq.~(\ref{drewriteimp})
in principle stands for the NLL Wilson coefficient, supplemented
by the $n_f$ dependent NNLL contributions. As the latter 
are not known yet, we take this Wilson coefficient at NLL precision
in the numerical evaluations. The Wilson coefficients entering
$V^{(1)}(\mu)$ are in principle the LL coefficients, supplemented
by the $n_f$ dependent NLL contributions. In practice, we decide
to replace these Wilson coefficients by the respective complete NLL version.
Finally, the Wilson coefficients 
entering $V^{(2)}(\mu)$ are the LL versions.
Note, that the gluon bremsstrahlung and the quark-antiquark emission
processes associated with $O_7$ are effectively transferred 
into $r_7^{(2)}$, $l_7^{(2)}$ and $t_7^{(2)}$, as described in 
Section~\ref{sec:o7calc}.
As already mentioned above, the square of the so-defined amplitude
incorporates the major part of the branching ratio. We therefore consider
the additional finite bremsstrahlung corrections to the decay width
only at the NLL level, i.e. we do not calculate  
the ${\cal O}(\as^2 n_f)$ corrections to these contributions.  

\begin{figure}[t]
  \begin{tabular}{cc}
  \includegraphics[bb=30 240 540 590,height=5.5cm]{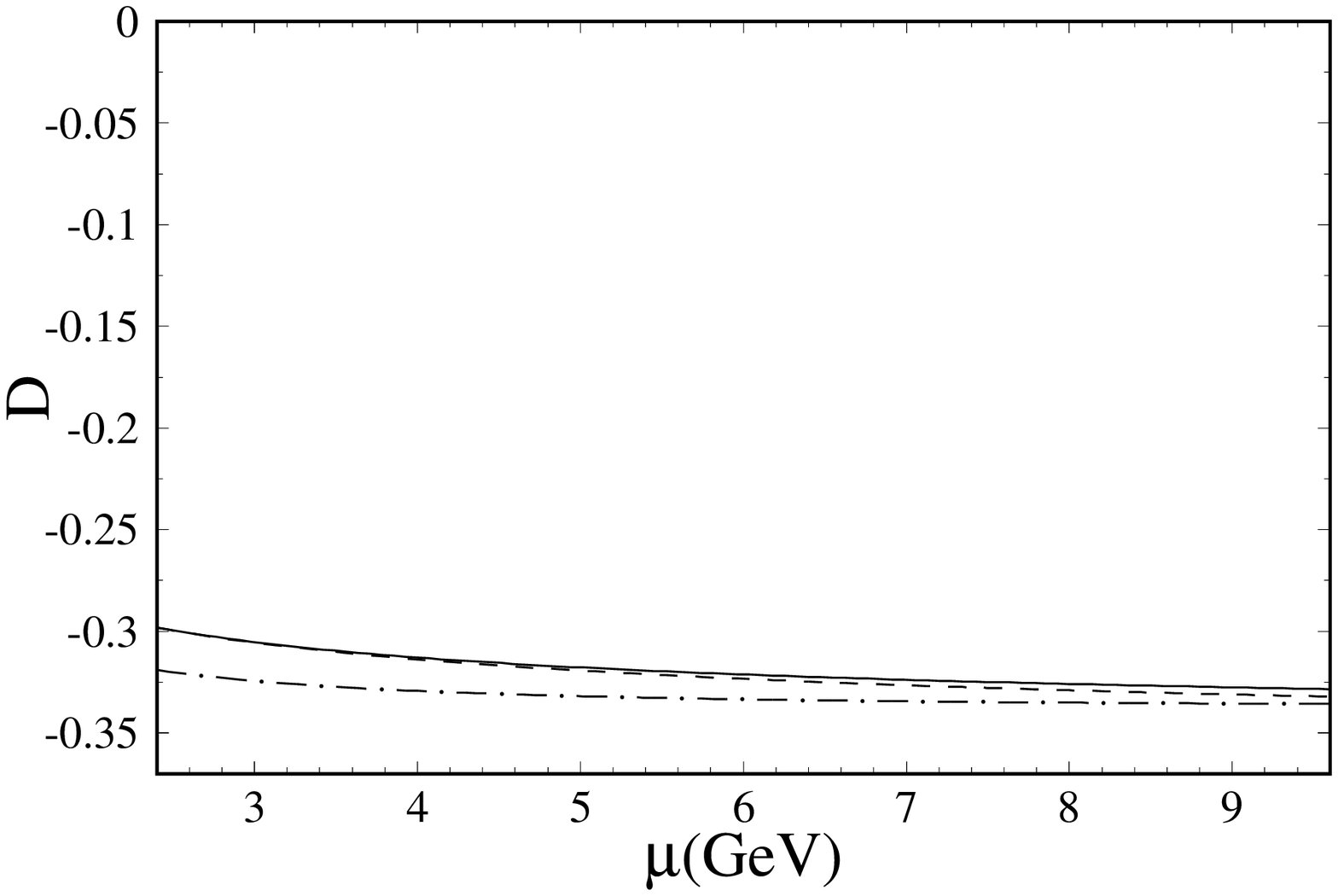}
  &
  \includegraphics[bb=30 240 540 590,height=5.5cm]{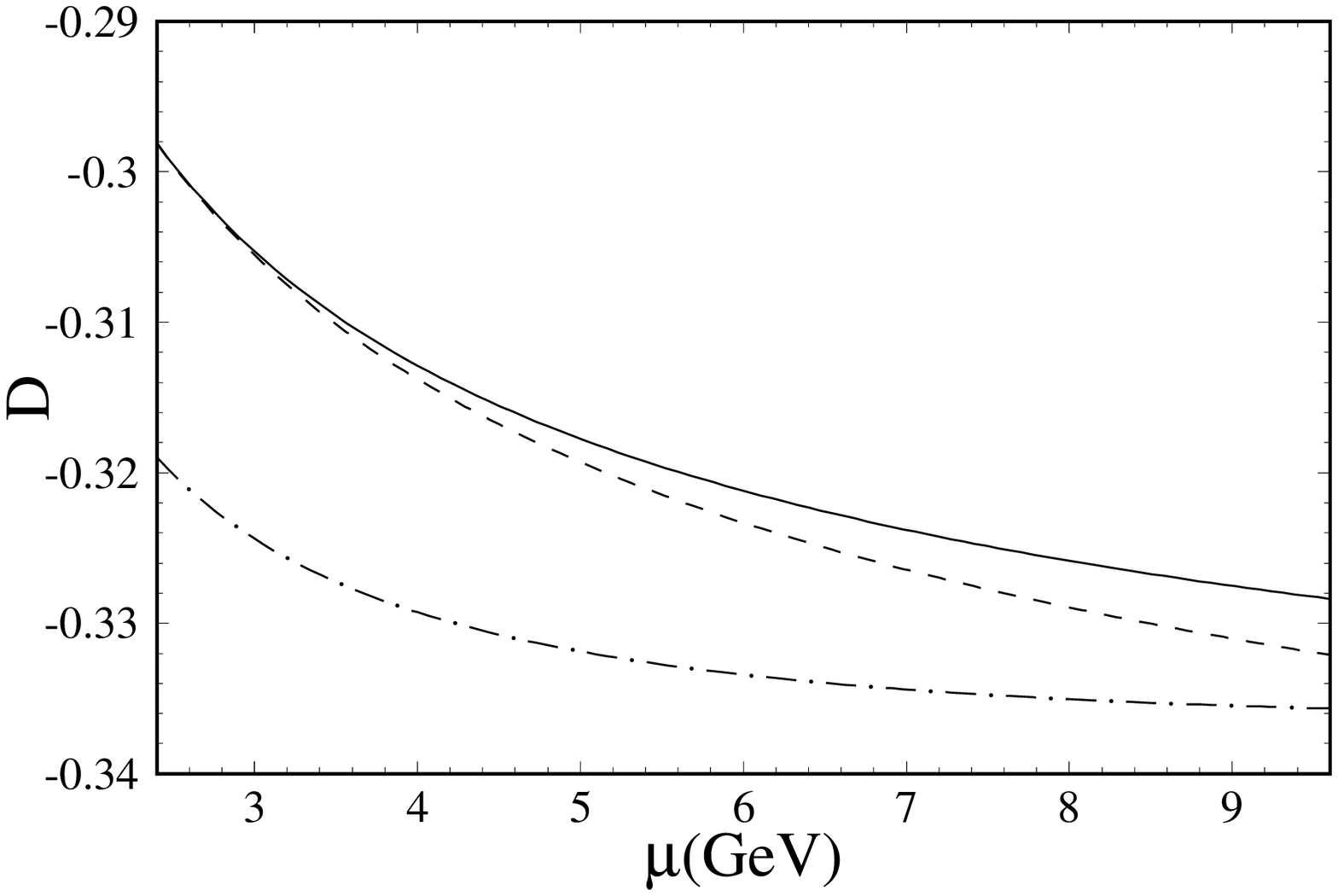}
  \end{tabular}
  \caption{\label{fig:amplitude}The reduced amplitude $D$ as a function of the
    renormalization scale $\mu$ where the 
    plot on the right is an enlargement of the one on the left.
    The dash-dotted curve represents the NLL approximation and the 
    solid curve includes the corrections of ${\cal O}(\alpha_s^2 n_f)$.
    For comparison we also show the result where the Wilson coefficients
    in $V^{(1)}$ (cf. Eq.~(\ref{vdefine})) are inserted to LL precision only
    (dashed curve).}
\end{figure}
As the square of the amplitude for $b \to s \gamma$ (in the sense defined
above) encodes the dominant part of the decay width,
it is reasonable to compare the NLL result 
$D^{{\rm NLL}}$ in Eq.~(\ref{drewrite})
with the corresponding 
${\cal O}(\as^2 n_f)$-improved result $D$
in Eq.~(\ref{drewriteimp}). In Fig.~\ref{fig:amplitude},
the function $D$ is plotted as a function of the renormalization
scale $\mu$.
We note, as already discussed in the Introduction, that we use in the
numerical evaluations
the hypothesis of naive non-abelianization, which amounts to
replacing $n_f$ by $-3 \beta_0/2$.
Nevertheless, in the following we still write 
${\cal O}(\alpha_s^2 n_f)$.
The dash-dotted line shows the NLL approximation as defined in 
Eq.~(\ref{drewrite}), while the solid curve shows the result after including
the ${\cal O}(\as^2 n_f)$ terms as discussed above. The dashed line
shows the result with ${\cal O}(\as^2 n_f)$ improvements,
in which, however, the Wilson coefficients in $V^{(1)}(\mu)$
are taken in LL approximation. The three curves illustrate that
the changes between the ${\cal O}(\as^2 n_f)$ improved version (solid line)
and the NLL prediction (dash-dotted line) are mainly due to the new
${\cal O}(\as^2 n_f)$ corrections of the matrix elements calculated in the
previous sections.

{}From ${\cal A}(b \to s \gamma)$ in Eq.~(\ref{amplitudeimp}) the 
decay width $\Gamma(b \to s \gamma)$ is easily obtained to be
\beq
\Gamma (b \to s \gamma) = \frac{G_F^2}{32\pi^4} \, 
 \vert \lamt\vert ^2 \alpha_{\rm em} \,
 m_b^5 \, |D|^2 \,.
\eeq
When giving numerical results for the NLL predictions, we only retain terms
in $|D|^2$ up to order $\as$, while for the improved version
we retain terms up to ${\cal O}(\as^2 n_f)$ in $|D|^2$
and systematically dismiss higher order contributions. For completeness
we should mention that $\alpha_s(\mu)$ is evaluated using two-loop 
accuracy in the $\beta$ function. We checked that 
the contribution of the three-loop term $\beta_2$ is numerically small.

To obtain the inclusive decay rate for $b \to X_s \gamma$, 
we have to take into account those terms 
which have not yet been absorbed into the virtual corrections.
At NLL precision, these contributions consist 
of those gluon bremsstrahlung corrections which are finite when
the gluon energy goes to zero; they have been calculated in 
Refs.~\cite{AG,POTT}. As the $(O_8,O_8)$
contribution to $\Gamma(b \to s \gamma g)$ becomes infrared singular
for {\it soft photon energies},
we introduce a photon energy cutoff $E_{\rm cut}$ as
in Ref.~\cite{Chetyrkin:1996vx}
and define the kinematical decay width
\beq
\Gamma(b \to X_s \gamma)_{E_\gamma \ge E_{\rm cut}} \, .
\eeq
At NLL the gluon bremsstrahlung contribution to this quantity can be written as
\beq
\label{brems}
\Gamma(b \to s \gamma g)_{E_\gamma \ge E_{\rm cut}}= 
\frac{G_F^2}{32 \pi^4} \, 
 \vert \lamt \vert ^2 
 \alpha_{\rm em} \, m_b^5 \, A \,,
\eeq
where $A$ is of the form \cite{Chetyrkin:1996vx}
\beq
\label{aterm}
A = \left( e^{-\as(\mu) \ln(\delta) (7+2\ln(\delta))/(3\pi)}-1\right) +
\frac{\as(\mu)}{\pi} \, \sum_{i,j=1;i \le j}^8 \, 
  {\rm Re} \left[ C_i^{{\rm eff}}(\mu) \,
            C_j^{{\rm eff}}(\mu)
 \, f_{ij}(\delta) \right] \, .
\eeq
The quantity $\delta$ is defined through
\beq
  E_{\rm cut} = \frac{m_b}{2} (1-\d) = E_{\rm max} (1-\d)\, .
\eeq
In \Eref{aterm} we put $C_{i}^{{\rm eff}}=0$ for 
$i=3,\ldots,6$, as in the virtual contributions.
We list the explicit expressions for the quantities $f_{ij}(\delta)$
in Appendix \ref{sec:rili}.

We should repeat that the ${\cal O}(\as^2 n_f)$ corrections are incorporated
in the quantity $D$, defined in Eqs.~(\ref{amplitudeimp}) and
(\ref{drewriteimp}).  We stress that the absorbed gluon bremsstrahlung- and
the quark-pair emission terms were obtained by integrating over the full range
of the photon energy.  Thus, since
we decided to implement a photon energy cut as just
described, the final expression for the kinematical decay width can be written
as
\beq
\label{widthrad}
\Gamma(b \to X_s \gamma)_{E_\gamma \ge E_{\rm cut}} \ =  
\frac{G_F^2}{32\pi^4}  
\vert \lamt \vert ^2 
\alpha_{\rm em} \, m_b^5 \, (|D|^2 +A) -
\Gamma_{77}^{(2),n_f}(b \to X_s \gamma)_{E_\gamma \le E_{\rm cut}}
 \, ,
\eeq 
where the expression for 
$\Gamma_{77}^{(2),n_f}(b \to X_s \gamma)_{E_\gamma \le E_{\rm cut}}$
is derived in Appendix \ref{sec:cutoff}.

In a last step, the kinematical branching ratio is obtained as
\beq
\label{br}
{\rm BR}(b \to X_s \gamma)_{E_\gamma \ge E_{\rm cut}} =  
\frac{\Gamma(b \to X_s \gamma)_{E_\gamma \ge E_{\rm cut}}}{\Gamma_{{\rm SL}}} 
 \, {\rm BR}_{{\rm SL}}\,,
\eeq
where ${\rm BR}_{{\rm SL}}$ is the measured semileptonic branching
ratio and the  
semileptonic decay width $\Gamma_{{\rm SL}}$ (supplemented
by the ${\cal O}(\as^2 n_f)$ terms~\cite{Luke:1994yc})
is given by ($z=m_c^2/m_b^2$)
\beq
\label{semileptonic}
\Gamma_{{\rm SL}} =
\frac{G_F^2|V_{cb}|^2 m_b^5}{192\pi^3} \, g(z) 
\left[ 1 - \frac{2 \as(\mu)}{3\pi}  \frac{h(z)}{g(z)} - 
  \left(\frac{\as(\mu)}{\pi}\right)^2  \beta_0  
  \left( \chi_{\beta}\left(\frac{m_c}{m_b}\right) 
    - \frac{1}{3} \frac{h(z)}{g(z)} 
    \ln \left( \frac{m_b}{\mu} \right)
  \right) \right] \,,
\eeq 
where the phase space function $g(z)$ and the ${\cal O}(\as)$
radiation function $h(z)$~\cite{Nir}
read
\begin{eqnarray}
  \label{gfun}
  g(z) &=&1-8z+8z^3-z^4-12z^2 \ln (z) \,,
  \nonumber\\
  \label{hfun}
  h(z) &=&\mbox{} -(1-z^2) \, \left( \frac{25}{4} - \frac{239}{3} \, z +
    \frac{25}{4} \, z^2 \right) + z \, \ln(z) \left( 20 + 90 \, z
    -\frac{4}{3} \, z^2 + \frac{17}{3} \, z^3 \right)
  \nonumber \\
  &&\mbox{} + z^2 \, \ln^2(z) \, (36+z^2)
  + (1-z^2) \, \left(\frac{17}{3} -\frac{64}{3} \, z + \frac{17}{3} \, z^2
  \right) \, \ln (1-z) \nonumber \\
  &&\mbox{} -4 \, (1+30 \, z^2 + z^4) \, \ln(z) \ln(1-z)
  -(1+16 \, z^2 +z^4)  \left( 6 \, \mbox{Li}_2(z) - \pi^2 \right)
  \nonumber \\
  &&\mbox{} -32 \, z^{3/2} (1+z) \left[\pi^2 - 4 \, \mbox{Li}_2(\sqrt{z})+
    4 \, \mbox{Li}_2(-\sqrt{z}) - 2 \ln(z) \, \ln \left(
      \frac{1-\sqrt{z}}{1+\sqrt{z}} \right) \right] \, .
\end{eqnarray}
The function $\chi_{\beta}(m_c/m_b)$, which encodes the 
${\cal O}(\as^2 n_f)$ 
terms\footnote{Note, that $n_f$ is replaced by $-3\beta_0/2$.}
is given in the form of a plot in Ref.~\cite{Luke:1994yc}. For 
$m_c/m_b=0.29$, which is the default value in our paper,
one finds $\chi_{\beta}(0.29) \approx 1.68$.

\begin{figure}[t]
  \begin{tabular}{cc}
  \includegraphics[bb=30 190 580 590,height=5.5cm]{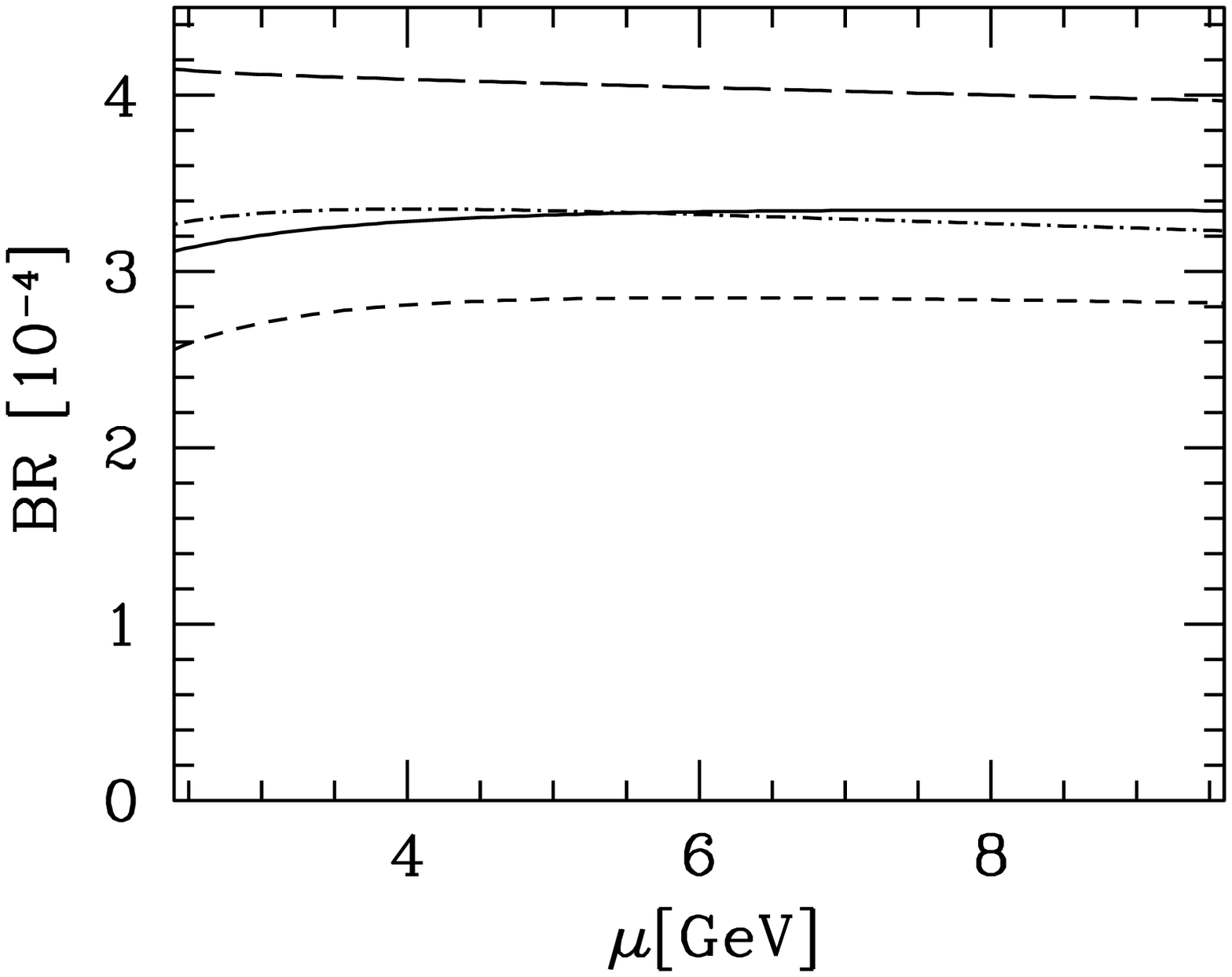}
  &
  \includegraphics[bb=30 190 580 590,height=5.5cm]{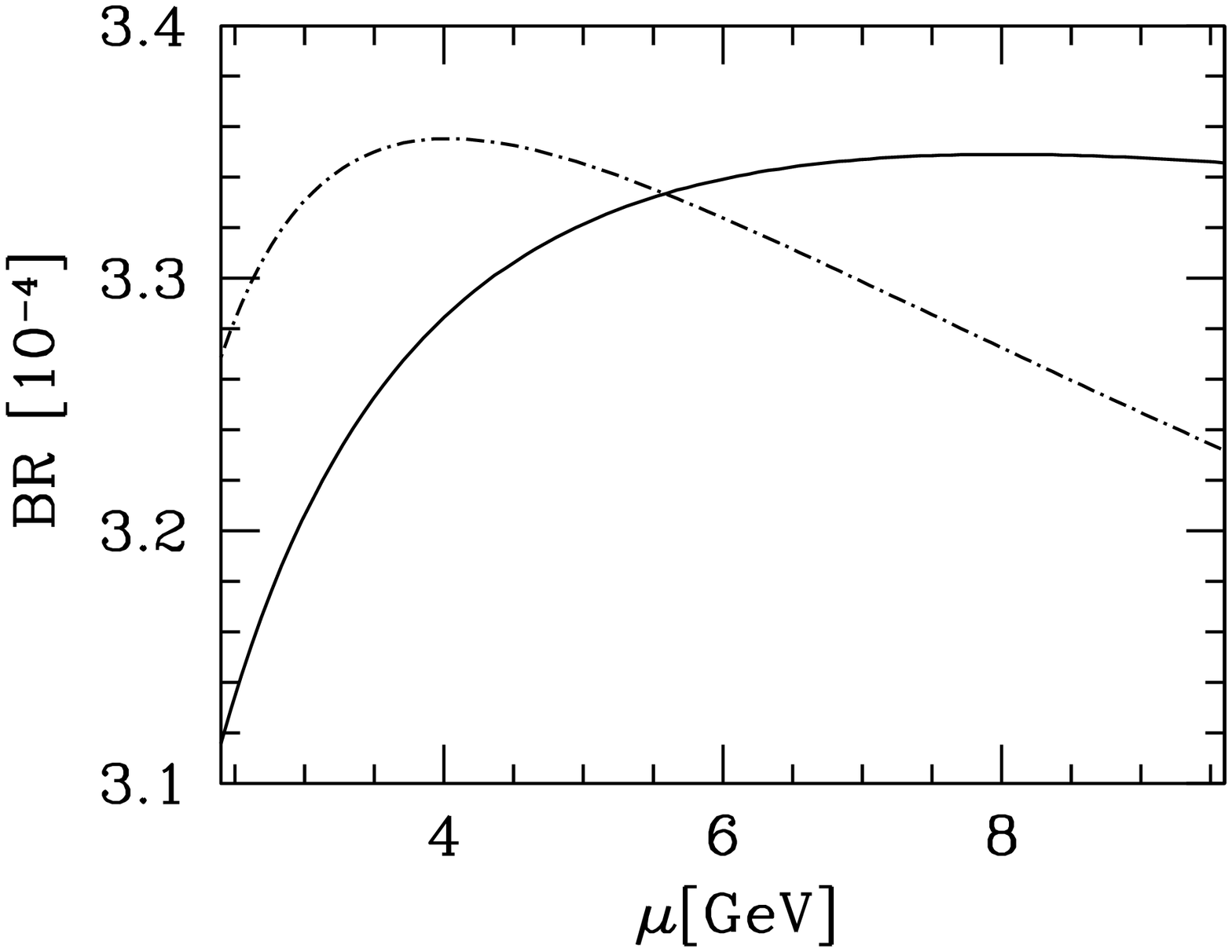}
  \end{tabular}
  \caption{\label{fig:branching}The branching ratio as a function of the
    renormalization scale $\mu$ where the 
    plot on the right is an enlargement of the one on the left.
    The dash-dotted curve represents the NLL approximation and the 
    solid curve includes the corrections of ${\cal O}(\alpha_s^2 n_f)$.
    For illustration in the left plot
    the latter are also shown for the case where
    $M_{1/2}^{(2)}$ ($M_7^{(2)}$) is set to zero which corresponds
    to short-dashed (long-dashed) curve. A photon energy cut of
  $E_{\rm cut}=m_b/20$ is used, which corresponds to $\delta=0.9$.}
\end{figure}
In Fig.~\ref{fig:branching} the kinematical branching ratio is shown 
for the choice
$E_{\rm cut} = m_b/20$, or, equivalently,
$\delta=0.9$~\cite{Kagan:1999ym} 
as a function of the
renormalization scale $\mu$. 
The input parameters were chosen to be:
$m_b=4.8$ GeV, $m_c/m_b=0.29$,
$m_t=173.8$ GeV,
$m_W=80.41$ GeV,
$m_Z=91.187$ GeV, $\as(m_Z)=0.119$, $\alpha_{{\rm em}}=1/137.036$,
$|\lamt /V_{cb}|^2=0.95$
and BR$_{\rm SL}=10.49\%$.
The dash-dotted line shows the branching ratio
${\rm BR}(b \to X_s \gamma)$ in NLL precision. In this case the
terms of ${\cal O}(\as^2 \beta_0)$ are consistently omitted in the 
expression for $\Gamma_{{\rm SL}}$ in Eq.~(\ref{semileptonic}).
The solid line shows the branching ratio where the ${\cal O}(\as^2 n_f)$
(or the ${\cal O}(\as^2 \beta_0)$) improvements are included. 

One observes that for $\mu\approx5.5$~GeV the ${\cal O}(\alpha_s^2 n_f)$
corrections vanish and that they are negative (positive) for smaller (larger)
values of $\mu$. In this context it is instructive to look at the
decomposition of the result. For this reason we show in the left plot of 
Fig.~\ref{fig:branching} the ${\cal O}(\alpha_s^2 n_f)$ corrections where
either $M_1^{(2)}$ and $M_2^{(2)}$ or $M_7^{(2)}$ 
is artificially set to zero which corresponds to the
short-dashed  and long-dashed curve, respectively.
This illustrates that there is a large cancelation between the 
negative contribution from $O_7$  and the one from
$O_1$ and $O_2$ which is, of course, also present in the amplitude $D$.
The effect of the $\alpha_s^2 n_f$ corrections from
the operator $O_8$ is significantly smaller and at
most of the order of 2\% in the considered interval for $\mu$.

Fig.~\ref{fig:branching} furthermore 
illustrates that the $\mu$ dependence of the
${\cal O}(\as^2 n_f)$ improved prediction for the branching ratio
is somewhat flatter than in the NLL case if we restrict ourselves to 
$\mu\ge4$~GeV. This is a welcome feature of our result, however,
in general we cannot expect to reduce the $\mu$ dependence as the solid curve
only represents a part of the ${\cal O}(\alpha_s^2)$ result.
Indeed, we obtain a stronger $\mu$-dependence in the region below 
$4$~GeV.

\begin{figure}[t]
  \includegraphics[height=7.9cm]{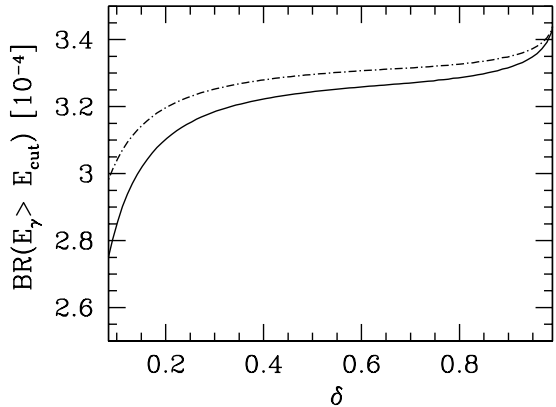}
  \caption{\label{fig:delta_dependence} Dependence of the branching
   ratio on the photon energy cut, 
   $E_{\rm cut}=\frac{m_b}{2}(1-\d)$. The dash-dotted curve shows the
   NLL result, while the solid curve includes the ${\cal O}(\a_s^2 n_f)$
   improvements. The renormalization scale is $\mu=4.8$ GeV.}
\end{figure}

In \Fref{fig:delta_dependence} we show the dependence of the
kinematical branching ratio on the photon energy cut. The dash-dotted line
shows the NLL result, while the solid curve includes the order
$\a_s^2 n_f$ improvements. We should mention at this point
that we did not include any non-perturbative effects in the photon
energy spectrum. The main purpose of this figure is to illustrate
how the order $\a_s^2 n_f$ contributions modify the NLL result. 

\section{\label{sec:concl}Conclusions}

In this paper a first step towards a complete NNLL calculation 
is undertaken
and radiative corrections to the matrix elements of
the operators $O_1$, $O_2$, $O_7$ and $O_8$ are computed.
More precisely, we consider the contributions of order $\alpha_s^2 n_f$
which are induced by a massless quark loop.
It is expected that these corrections, after replacing $n_f$ by
$-3\beta_0/2$, may give an important contribution to the full
order $\alpha_s^2$ corrections.
Furthermore, motivated by the NLL analysis, we expect that the
${\cal O}(\as^2 n_f)$ 
corrections to the matrix elements numerically dominate the ones
of the same order 
to the Wilson coefficient functions and to the anomalous dimension matrix.

In practice our calculation requires the evaluation of two- and 
three-loop diagrams in the case of $O_7$, $O_8$ and $O_1$, $O_2$,
respectively.
Furthermore, in order to obtain an infrared finite result in the case of $O_7$,
also the contributions from the gluon bremsstrahlung and from the
quark-pair emission process are taken into account which 
requires the evaluation of 
three- and four-particle phase space integrals, respectively.
All calculations are performed analytically where an expansion in 
$m_c/m_b$ is applied to the three-loop diagrams.
For practical purposes this expansion is equivalent to the exact result.

As far as the numerical impact of our result is concerned, we observe a
striking cancelation among the individual contributions at order $\alpha_s^2
n_f$.  When using a photon energy cut of $E_{\rm cut}=m_b/20$, the ${\cal
O}(\as^2 n_f)$ terms reduce (after replacing $n_f$ by $-3\beta_0/2$) 
the branching ratio by $-0.98\%$ for
$\mu=m_b=4.8$~GeV and lead to corrections of $-3.9\%$ and $+3.4\%$ for
$\mu=3.0$~GeV and $\mu=9.6$~GeV, respectively.

\section*{Acknowledgements}
We would like to thank M. Misiak for making available to us his results for
the renormalization constants of the operator mixing which provided important
checks for our calculation.  M.S. thanks T.~Teubner for discussions on the
quark-pair emission process.  We thank A.~Parkhomenko for carefully reading
the manuscript.  Our work is partially supported by the Swiss National
Foundation and by RTN, BBW-Contract N0. 01.0357 and EC-Contract
HPRN-CT-2002-00311 (EURIDICE).

\appendix


\section{\label{sec:build}Building Blocks}

\begin{figure}[t]
  \begin{center}
    \leavevmode
    \includegraphics[height=4.2cm,angle=0]{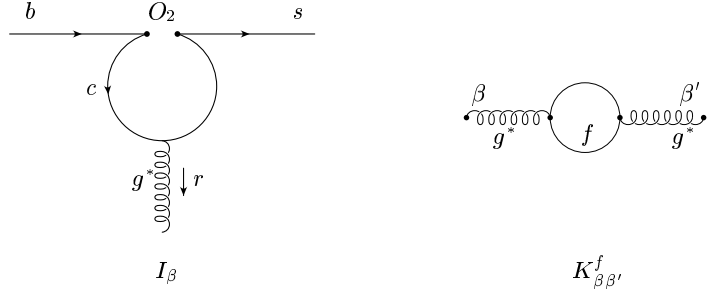}
  \end{center}
  \caption{\label{fig:build1} The building blocks $I_{\b}$ and $K_{\b\b'}^f$
    which are used in the calculation of the Feynman diagrams. The curly lines
    represent virtual gluons, whereas the letters $b$, $c$ and $s$ stand for
    the corresponding quark ($f$ stands for a generic quark of mass $m_f$). 
    Note that the
    external gluons are not amputated in the case of $K_{\b\b'}^f$.}
\end{figure}

\begin{figure}[t]
  \begin{center}
    \leavevmode \includegraphics[height=4cm]{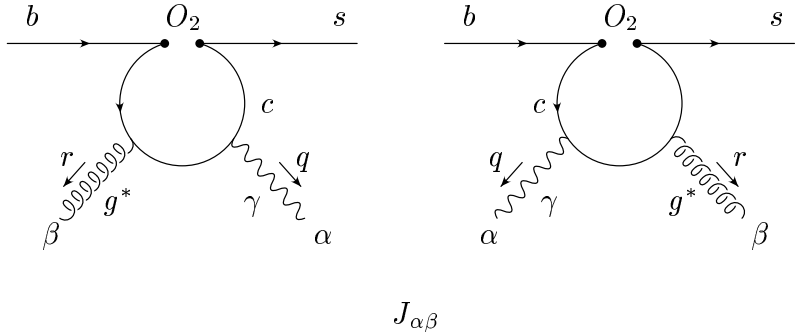}
  \end{center}
  \caption{\label{fig:build2} The building block $J_{\a \b}$ used in the
    calculation of the Feynman diagrams involving $O_1$ and $O_2$. 
    The curly and wavy
    lines represent off-shell gluons and on-shell photons, respectively.}
\end{figure}

The three-loop diagrams involving $O_1$ and $O_2$ as well as the two-loop
graphs involving $O_7$ and $O_8$ can be calculated by using one or more of the
building blocks $I_\b$, $J_{\a\b}$ and $K_{\b\b'}^f$ to be discussed in this
appendix. The corresponding
diagrams are shown in Figs.~\ref{fig:build1} and~\ref{fig:build2}
where the color indices are suppressed. 

The calculation of $I_{\b}$ is straightforward and yields
\bea
  \label{eq:build1}
  I_\beta &=& - \frac{g_s}{4\, \pi^2} \, \Gamma(\epsilon) \, \mu^{2 \epsilon}
  \, e^{\gamma_E \epsilon} \, (1-\epsilon) \, e^{i \pi \epsilon} \,
  \left(r_{\beta} \dslash{r} - r^2 \gamma_\beta \right) \, L \,
  \frac{\lambda}{2}  \nn \\
  && \int_0^1 \! {\rm d}x \,
  \left[x(1-x)\right]^{1-\epsilon} \, \left[ r^2 - \frac{m_c^2}{x(1-x)} + i\,
  \delta \right]^{-\epsilon} \, ,
\eea
where $r$ is the momentum of the virtual gluon emitted from the $c$-quark
loop. In the three-loop diagrams shown in \Fref{fig:M1M2} 
(cf. Section~\ref{sec:threeloop}), 
the free index $\b$ will be contracted with the corresponding index of the
dressed gluon propagator  $K_{\b\b'}^f$. 

It is also quite simple to obtain 
the building block $K_{\b \b'}^f$ (i.e., the dressed gluon propagator)
which can be cast into the form
\begin{equation}
  \label{eq:build3}
  K_{\b \b'}^f = -\frac{g_s^2}{2 \p^2} T \, \Gamma(\e) e^{\g_E \e} e^{i\p\e}
  \m^{2\e}\frac{1}{i} \frac{g_{\b \b'}-\tfrac{r_\b
      r\raisebox{-0.28ex}{$\scriptscriptstyle \b'$}}{r^2}}{r^2+i\delta}
  \int_0^1\,{\rm d}x\,
  x(1-x) \left(x(1-x)r^2-m_f^2+i\d\right)^{-\e},
\end{equation}
where $m_f$ denotes the mass of the quarks and $T=\frac{1}{2}$. Note that
this expression is independent of the gauge parameter $\xi$ which enters
the free gluon propagators in the construction of $K_{\b \b'}^f$,
when working in an arbitrary $R_{\xi}$ gauge. 

The building block $J_{\a \b}$ is somewhat  more involved. Adopting the 
notation of Ref.~\cite{Simma:1990nr}, it reads (for an on-shell photon)
~\cite{Greub:1996tg}
\begin{eqnarray}
\label{eq:bb2}
    J_{\alpha\beta} &=& \frac{e\,g_s\,Q_u}{16\,\pi^2} \left[ E(\alpha,\beta,r)
    \Delta i_5 + E(\alpha,\beta,q) \Delta i_6 -
    E(\beta,r,q)\frac{r_{\alpha}}{\qr}\, \Delta i_{23}
    \right. \nn \\ &&\left.  -
    E(\alpha,r,q)\frac{r_{\beta}}{\qr}\, \Delta i_{25} -
    E(\alpha,r,q)\frac{q_{\beta}}{\qr}\, \Delta i_{26} \right] \, L \,
    \frac{\lambda}{2} \, ,
\end{eqnarray}
where $q$ and $r$ denote the momenta of the on-shell photon and the off-shell
gluon, respectively. 
When inserted into the full diagrams in Fig.~\ref{fig:M3M4},
the indices $\alpha$ and $\beta$ will be contracted with the
polarization vector $\varepsilon$ of the photon and with
the dressed gluon propagator $K_{\b \b'}^f$, respectively. The matrix
$E(\alpha,\beta,r)$ is defined as
\begin{equation}
    E(\alpha,\beta,r) = \frac{1}{2} (\gamma_{\alpha}\gamma_{\beta}\dslash{r} -
    \dslash{r}\gamma_{\beta}\gamma_{\alpha}),
\end{equation}
and the dimensionally regularized quantities $\Delta i_k$ occurring in
\Eref{eq:bb2} read
\begin{eqnarray}
  \Delta i_5 &=& 4\, B^+ \int_S\! {\rm d}x\,{\rm d}y 
  \left[ 4 (\qr)\, x\, y \,(1-x)
    \epsilon + r^2 \,x\, (1-x) (1-2\,x) \epsilon \right. \nonumber \\ &&
  \hspace{1cm} \left. +
    (1-3x)C \right] \, C^{-1-\epsilon} \, , \nonumber \\ \nonumber \\ \Delta
  i_6 &=& 4\, B^+ \int_S\! {\rm d}x\,{\rm d}y 
  \left[ -4 (\qr)\, x\, y \,(1-y) \epsilon
  -r^2\, x\, (2 - 2\,x + 2\,x\,y - y) \epsilon \right. \nonumber \\ &&
  \hspace{1cm} \left.
    -(1-3\,y)C \right] \, C^{-1-\epsilon} \, , \nonumber \\ \nonumber \\
  \Delta i_{23} &=& - \Delta i_{26} = 8\, B^+ (\qr) \int_S\! 
  {\rm d}x\,{\rm d}y\,
  x\, y \,\epsilon \, C^{-1-\epsilon} \, , \nonumber \\ \nonumber \\
  \Delta i_{25} &=& -8\, B^+ (\qr) \int_S\! {\rm d}x\, {\rm d}y\, x\, (1-x)
  \,\epsilon \, C^{-1-\epsilon} \, ,
\end{eqnarray}
where $B^+ = (1+\epsilon) \Gamma(\epsilon)\, e^{\gamma_E \epsilon}
\mu^{2\epsilon}$ and $C$ is given by
\begin{equation*}
  C = m_c^2 - 2\, x\, y (\qr) - r^2\, x\, (1-x) - i \delta.
\end{equation*}
The integration over the Feynman parameters $x$ and $y$ is restricted to the
simplex $S$, i.e. $y\in[0,1-x]$, $x\in[0,1]$. Due to Ward identities, the
quantities $\Delta i_k$ are not independent of one another. Namely,
\[q^\alpha J_{\alpha\beta} = 0 \quad \text{and} \quad r^\beta
J_{\alpha\beta} = 0 \]
imply that $\Delta i_5$ and $\Delta i_6$ can be expressed as
\begin{equation}
    \Delta i_5 = \Delta i_{23} \,, \quad
    \Delta i_6 = \frac{r^2}{\qr}\, \Delta i_{25} + \Delta i_{26}.
\end{equation}

\section{Regularized three-loop results 
  for $\bm{\langle s \gamma|O_2|b \rangle}$}
\label{sec:threeloopresults}
In Section~\ref{sec:threeloop} 
we explained in some detail the calculation of the
virtual three-loop corrections to $\langle s \gamma|O_2|b\rangle$. 
Here we give the results for the
four gauge-invariant sets of graphs depicted in Figs.~\ref{fig:M1M2} and
~\ref{fig:M3M4}. The results read, using $z = m_c^2/m_b^2$ and $L = \ln(z)$: 
\bea 
M_{2,{\rm bare}}^{(2)}(1) 
& = & \Bigg\{ \frac{1}{\e}\Bigg[ - \frac{1}{81\e} - \frac{29}{243} +
\frac{1}{6}\left(5 + 2L\right)z + \frac{1}{6}\left(5 - 2L + 2L^2 -
  2\p^2\right)z^2\nn \\ && +\frac{1}{81}\left(17 + 30L - 18L^2 +
  18\p^2\right)z^3 - \frac{i\p}{27}\left(1 - 9z + 9z^2 - 18Lz^2\right.\nn\\ &&
\left.- 10z^3 + 12Lz^3\right) \Bigg] \left( \frac{m_b}{\m}\right)^{-6\e} +
\Bigg[-\frac{1063}{1458} + \frac{19\p^2}{324} \nn \\ && + \frac{1}{18} \left(61
  + 4L - 9L^2 - 10\p^2\right)z + \frac{1}{18}\left(79 - 22L + 28L^2 - 8L^3 -
  9\p^2 \nn \right. \\ && \left. -14L\p^2 - 12 \zeta(3) \right)z^2 +
\frac{1}{81}\left(63 - 27L - 36L^2 + 24L^3 - 59\p^2 + 42L\p^2 + \right. \nn \\
&& \left. 36\zeta(3)\right)z^3 - \frac{i\p}{162}\Big(58 - 441z - 9\left(23 +
  38L - 6L^2 - 12\p^2 \right)z^2 \nonumber \\ && -12\left(4 + 3L + 3L^2 +
  6\p^2 \right) z^3 \Big) \Bigg] + {\cal O}(z^4)\Bigg\}  \left(\frac{\a_s}{\p}\right)^2 C_F T
n_f Q_d \bra s \g | O_7 | b \ket_{\text{tree}}, \nn \\ &&\\
M_{2,{\rm bare}}^{(2)}(2) 
&=& \Bigg\{ \frac{1}{\e}\Bigg[\frac{7}{162\e} + \frac{5}{486} +
\frac{1}{18} \left(3 - \p^2\right)z + \frac{2\p^2}{9}z^{3/2} -
\frac{1}{6}\left(6 - 6L + L^2\right)z^2\nn\\ && +\frac{1}{324}\left(157 - 6L -
144L^2 - 60\p^2\right)z^3\Bigg] \left(\frac{m_b}{\m}\right)^{-6\e}\nn \\ && +
\Bigg[-\frac{1387}{1458} + \frac{11\p^2}{72} + \frac{1}{54}\left(96 - 17\p^2 -
126\zeta(3)\right)z\nn\\ && +\frac{\p^2}{27}\left(40 - 18L -
72\ln(2)\right)z^{3/2} \nn \\ && +\frac{1}{36}\left(213 + 102L - 40L^2 + 8L^3
+ 34\p^2 + 96\zeta(3)\right)z^2 - \frac{20\p^2}{9}z^{5/2}\nn\\ &&
+\frac{1}{324} \left(2799 - 995L - 198L^2 + 192L^3 - 10\p^2 - 60L\p^2 -
936\zeta(3) \right)z^3\Bigg] \nn \\ && + {\cal O}(z^{7/2})\Bigg\}
\left(\frac{\a_s}{\p}\right)^2 C_F T n_f Q_d \bra s \g | O_7 | b
\ket_{\text{tree}},
\eea
\bea 
M_{2,{\rm bare}}^{(2)}(3) 
&=& \Bigg\{\frac{1}{\e}\Bigg[\frac{1}{36\e} +
\frac{137}{432} - \frac{1}{36}\left(18 + 24L + 3L^2 + 2L^3 - 3\p^2 - 6L\p^2 -
  24\zeta(3)\right)z\nn\\ && -\frac{1}{36}\left(15 + 6L - 6L^2 + 2L^3 + 6\p^2
  - 6L\p^2 - 24\zeta(3)\right)z^2\nn\\ && +\frac{1}{36}\left(17 -
  12L\right)z^3 + \frac{i\p}{36}\left(3 - 24z - 6Lz -6L^2z + 2\p^2z - 6z^2 +
  12L z^2\right.\nn\\ && \left.-6L^2z^2 + 2\p^2z^2 - 12 z^3 \right) \Bigg]
\left(\frac{m_b}{\m}\right)^{-6\e} \nn \\ && +\Bigg[ \frac{6029}{2592} -
\frac{17\p^2}{144} - \frac{1}{1080}\left(7200 + 6240L - 120L^2 + 220L^3 -
  105L^4\right.\nn\\ && \left. -2040\p^2 - 1200L\p^2 + 90L^2\p^2 + 111\p^4 -
  4440\zeta(3) + 1440L\zeta(3)\right)z\nn\\ && -\frac{1}{2160}\left(15135 -
  5790L - 1050L^2 + 980L^3 - 210L^4 - 30\p^2 - 780L\p^2\right.\nn\\
&&\left. +180L^2\p^2 + 222\p^4 - 4560\zeta(3) + 2880L\zeta(3)\right)z^2 +
\frac{1}{72}\left(3 - 2L + 72\p^2\right)z^3\nn\\ && +\frac{i\p}{432}\Big(411 -
4\left(786 + 192L + 93L^2 - 24L^3 - 49\p^2 - 12L\p^2 - 72\zeta(3)\right)z\nn\\
&& +2\left(309 + 102L - 186L^2 + 48L^3 - 10\p^2 + 24L\p^2 +
  144\zeta(3)\right)z^2\nn\\ && +8\left(75 - 54L\right)z^3\Big) \Bigg] + {\cal
  O}(z^4)\Bigg\}
\left(\frac{\a_s}{\p}\right)^2C_F T n_f Q_u \bra s \g | O_7 | b
\ket_{\text{tree}},\\ 
M_{2,{\rm bare}}^{(2)}(4) &=& \Bigg\{ \frac{1}{\e}\Bigg[\frac{1}{18\e} +
\frac{127}{432} - \frac{1}{36}\left(12 + 6L - L^3 - \p^2 - 3L\p^2 -
12\zeta(3)\right)z\nn\\ && -\frac{1}{36}\left(6 - 6L + 3L^2 - L^3 + 2\p^2 +
24\zeta(3)\right)z^2 - \frac{1}{324}\left(27 + 108L\right.\nn\\ &&
\left. -81L^2 - 27\p^2\right)z^3\Bigg] \left(\frac{m_b}{\m}\right)^{-6\e} \nn
\\ && + \Bigg[ \frac{2839}{2592} + \frac{13\p^2}{144} -
\frac{1}{2160}\left(9480 + 2040L + 180L^2 - 340L^3 + 105L^4\right.\nn\\ &&
\left. + 260\p^2 - 720L\p^2 + 30L^2\p^2 - 439\p^4 - 3360\zeta(3) -
8640L\zeta(3)\right)z\nn\\ && -\frac{8\p^2}{3}z^{3/2} +
\frac{1}{4320}\left(29895 - 6270L - 1410L^2 + 740L^3 - 210L^4 +
920\p^2\right.\nn\\ &&\left. -480L\p^2 + 120L^2\p^2 - 52\p^4 - 16320\zeta(3) +
4320L\zeta(3)\right)z^2 + \frac{40\p^2}{27}z^{5/2}\nn\\ && -\frac{1}{216}
\left(1358 - 477L - 99L^2 + 90L^3 + 63\p^2 - 18L\p^2 -
432\zeta(3)\right)z^3\Bigg] \nn \\ && +{\cal O}(z^{7/2}) \Bigg\}
\left(\frac{\a_s}{\p}\right)^2C_F T n_f Q_u \bra s \g | O_7 | b
\ket_{\text{tree}}.
\eea
In these expressions, $\zeta$ denotes the Riemann $\zeta$ function with the
value $\zeta(3)\approx1.2020569$. $Q_u=2/3$
and $Q_d=-1/3$ are the electric charge factors of the up- and down-type quarks,
respectively, while $C_F=4/3$ and $T=1/2$ are color factors.

\section{Correction functions needed for the NLL result}
\label{sec:rili}

The renormalization scale independent parts of the virtual corrections in NLL
order precision, encoded in the functions $r_i^{(1)}$, appearing in
Eq.~(\ref{vdefine}), read
\bea
\label{ris}
 r_1^{(1)} & = & -\frac{1}{6} r_2^{(1)}\,,    \nonumber \\ 
 r_2^{(1)} & = &
  \frac{2}{243} \,\left\{-833+144 \pi^2 z^{3/2}
               \right.                  \nonumber \\
     &   &  \hspace{1.2cm}
+ \left[ 1728 -180 \pi^2 -1296 \,\zeta (3)
       +(1296 -324 \pi^2) L +108 L^2 +36 L^3
  \right]  z                            \nonumber \\
     &   &  \hspace{1.2cm}
+ \left[ 648 +72 \pi^2 +(432 -216 \pi^2) L +36 L^3
  \right]    z^2                        \nonumber \\
     &   &  \hspace{1.2cm} 
 \left.                 +
 \left[ -54 -84 \pi^2 +1092 L -756 L^2
 \right]    z^3  \,
 \right\}                               \nonumber \\
     &   &
 + \frac{16 \pi i}{81} \, \left\{ -5 
 + \left[ 45 -3 \pi^2 + 9 L + 9 L^2 \right]    z
 + \left[ -3 \pi^2 + 9 L^2 \right]    z^2
 + \left[ 28 - 12 L  \right]  z^3 \, \right\}  \nonumber \\
     &   &  \hspace*{1.2cm}
\hspace{0.3cm} + \hspace{0.3cm} {\cal O}(z^{7/2})\,,  \nonumber \\
 r_7^{(1)} & = &
   \frac{32}{9} -\frac{8}{9} \pi^2\,,             \nonumber \\
 r_8^{(1)} & = &
 -\frac{4}{27} ( -33 + 2 \pi^2 - 6 i \pi )\,,
\eea
where $z$ is defined as $z=m_c^2/m_b^2$ and the symbol
$L$ denotes $L=\ln(z)$. 
The quantities $l_i^{(1)}$, appearing  
in Eq.~(\ref{vdefine}), read
\beq
l_1^{(1)} = - \frac{1}{6} \, l_2^{(1)}\,,\qquad
l_2^{(1)} = \frac{416}{81}\,,\qquad
l_7^{(1)} = \frac{8}{3}\,,\qquad
l_8^{(1)} = - \frac{32}{9} \, .
\eeq
Notice that $r_3^{(1)}$, $r_4^{(1)}$, $r_5^{(1)}$ and $r_6^{(1)}$, as well
as 
$l_3^{(1)}$, $l_4^{(1)}$, $l_5^{(1)}$ and $l_6^{(1)}$ 
are  not needed 
in the approximation $C_i^{{\rm eff}}(\mu)=0$ ($i=3,4,5,6$).

The functions $f_{ij}$ needed for \Eref{aterm} are taken from
Ref.~\cite{Chetyrkin:1996vx} and are listed here for
completeness. Note that $f_{77}(\d)$ differs from the one given in
Ref.~\cite{Chetyrkin:1996vx} in order to be compatible with our $r_7$ given in
\Eref{ris}\footnote{The additional, $\d$-independent addend appearing in our
  $f_{77}(\d)$ is such that $f_{77}(1)$ vanishes: the contribution of
  $f_{77}(\d)$ at $\d=1$ is already absorbed into our $r_7$.}.
\begin{equation*}
  f_{11}(\d)=\tfrac{1}{36}f_{22}(\d)\,,\;\;
  f_{12}(\d)=-\tfrac{1}{3}f_{22}(\d)\,,\;\;
  f_{17}(\d)=-\tfrac{1}{6}f_{27}(\d)\,,\;\;
  f_{18}(\d)=-\tfrac{1}{6}f_{28}(\d)\,,
\end{equation*}
\begin{eqnarray}
  \label{fijs}
  f_{22}(\d) &=& \frac{16z}{27}\Bigg[\d\int_0^{(1-\d)/z} dt\,
  (1-zt)\left|\frac{G(t)}{t} +\frac{1}{2}\right|^2+\int_{(1-\d)/z}^{1/z} dt\,
  (1-zt)^2\left|\frac{G(t)}{t} +\frac{1}{2}\right|^2\Bigg]\,,\nn\\
  f_{27}(\d) &=& -\frac{8z^2}{9}\Bigg[\d\int_0^{(1-\d)/z} dt\,
  \mbox{Re}\left(G(t) +\frac{t}{2}\right) +\int_{(1-\d)/z}^{1/z} dt\, (1-zt)
  \mbox{Re} \left(G(t) +\frac{t}{2}\right)\Bigg]\,,\nn\\
  f_{28}(\d) &=& -\frac{1}{3}f_{27}(\d)\,,\nn\\
  f_{77}(\d) &=& \frac{10}{3}\d +\frac{1}{3}\d^2 -\frac{2}{9}\d^3
  +\frac{1}{3}\d(\d-4)\ln(\d) -\frac{31}{9}\,,\nn\\
  f_{78}(\d) &=& \frac{8}{9} \Bigg[\mbox{Li}_2(1-\d) -\frac{\p^2}{6} -\d\ln(\d)
  +\frac{9}{4} \d -\frac{1}{4}\d^2 +\frac{1}{12}\d^3\Bigg]\,,\nn\\
  f_{88}(\d) &=& \frac{1}{27}\Bigg\{ -2 \ln\left(\frac{m_b}{m_s}\right)\Big[
  \d^2 +2\d +4\ln(1-\d)\Big] + 4\mbox{Li}_2(1-\d) -\frac{2\p^2}{3}\nn\\ &&
  -\d(2+\d)\ln(\d) +8\ln(1-\d) +7\d +3\d^2 -\frac{2}{3}\d^3\Bigg\},
\end{eqnarray}
where the function $G(t)$ is defined through
\begin{eqnarray}
  \label{eq:gt}
  G(t) &=& \begin{cases} -2\,\mbox{arctan}^2\left(\sqrt{\frac{t}{4-t}}\right) &
  \text{for } t<4 \\ -\frac{\p^2}{2}
  +2\ln^2\left(\frac{1}{2}(\sqrt{t}+\sqrt{t-4})\right) -2i\p\ln
  \left(\frac{1}{2}(\sqrt{t}+\sqrt{t-4})\right) & \text{for } t\geq4.
 \end{cases}
\end{eqnarray}
The functions $f_{ij}$ associated with the operators $O_3-O_6$ are not
needed in our approximation.\\
Note that in the numerics we set $m_s$ equal to zero in all terms except
$f_{88}(\d)$, where a value of $m_b/m_s=50$ is chosen.

\section{$\bm{{\cal O}(\as^2 n_f)}$ contributions to various $\bm{Z}$ factors}
\label{sec:zfactors}

In this appendix we give the results for the ${\cal O}(\as^2 n_f)$
contributions for various $Z$ factors entering the calculation
of the counterterm $M_{7,{\rm ct}_2}^{(2),(a)}$ in 
Eq.~(\ref{m7virtcounttwo}) in Section~\ref{sec:o7calc}. For the meaning
of the various terms, see the text after 
Eq.~(\ref{m7virtcounttwo}). The ${\cal O}(\as^2 n_f)$ contributions to the
relevant $Z$ factors read
\begin{eqnarray}
  \delta Z_{2,b}^{(2),n_f} &=& \left(\frac{\alpha_s}{\pi}\right)^2 \frac{C_F T
  n_f }{288} \bigg(\frac{18}{\epsilon}\left(1 - 4\ln(f) -
  8\ln\left(\frac{m_b}{\mu}\right)\right) + 443\nonumber \\ && +30\pi^2+
  96\ln(f)+ 72\ln^2(f) - 264\ln\left(\frac{m_b}{\mu}\right) \nonumber \\
  && +288\ln(f)\ln\left(\frac{m_b}{\mu}\right) + 432
  \ln^2\left(\frac{m_b}{\mu}\right)\bigg)\,,
\end{eqnarray}
\begin{eqnarray}
  \delta Z_{2,s}^{(2),n_f}&=& \left(\frac{\alpha_s}{\pi}\right)^2 \frac{C_F T
  n_f }{96} \bigg( \frac{6}{\epsilon} \left(1 - 4\ln(f) -
  8\ln\left(\frac{m_b}{\mu}\right)\right) -5 + 2\pi^2 \nonumber \\ &&-
  44\ln(f)+ 12\ln^2(f)+ 24\ln(f)\ln(r) - 88\ln\left(\frac{m_b}{\mu}\right)
  \nonumber \\
  && + 96\ln(f) \ln\left(\frac{m_b}{\mu}\right) +
  48\ln(r)\ln\left(\frac{m_b}{\mu}\right)+144
  \ln^2\left(\frac{m_b}{\mu}\right)\bigg)\,, \\
  \delta Z_{m_b}^{{\rm on},(2),n_f} &=&\left(\frac{\alpha_s}{\pi}\right)^2
  \frac{C_F T n_f }{96} \left(71+8\pi^2-104\ln\left(\frac{m_b}{\mu}\right)+48
  \ln^2\left(\frac{m_b}{\mu}\right) \right.\nonumber \\ &&
  \left. +\frac{10}{\epsilon}-\frac{12}{\epsilon^2}\right)\,,\\ \delta
  Z_{77}^{(2),n_f} &=& \left(\frac{\alpha_s}{\pi}\right)^2 \frac{C_F n_f T}
  {36\epsilon} \left(\frac{6}{\epsilon}-7\right)\,,
\end{eqnarray}
with $r = m_s^2/m_b^2$ and $f = m_f^2/m_b^2$.  


\section{Implementing the photon energy cut-off in the $\bm{{\cal O}(\as n_f)}$
terms}
\label{sec:cutoff}

In this appendix we provide the
formulas which are needed to calculate the ${\cal O}(\as^2 n_f)$ piece of the
kinematical branching ratio 
$\mbox{BR}(b \to X_s \gamma)_{E_\gamma \ge E_{\rm cut}}$,
where $E_{\rm cut}$ represents a cut-off on the photon energy. 
As can be seen from the structure of \Eref{widthrad}, 
this amounts to calculate 
$\Gamma_{77}^{(2),n_f}(b \to X_s \gamma)_{E_\gamma \le E_{\rm cut}}$, 
which is contained in the quantity $D$ of \Eref{widthrad}.

Note that
only the gluon bremsstrahlung- and the quark-pair emission
processes enter the calculation for 
$\Gamma_{77}^{(2),n_f}(b \to X_s \gamma)_{E_\gamma \leq E_{\rm cut}}$
as the photon energy in the virtual contributions
is concentrated at $m_b/2$.
The ${\cal O}(\alpha_s^2 n_f)$ contribution to 
$\Gamma_{77}^{(2),n_f}(b \to X_s \gamma)_{E_\gamma \leq E_{\rm cut}}$
can be written in the form
\beq
\Gamma^{(2),n_f}_{77}(b \to X_s \gamma)_{E_\gamma \leq E_{\rm cut}}=
\Gamma_{77}^0 \left[ \hat{\Gamma}_{77}^{(2),(b)}(E_\gamma \leq E_{\rm cut})+ 
                     \hat{\Gamma}_{77}^{(2),(c)}(E_\gamma \leq E_{\rm cut})
              \right] \, ,
\label{eq:gamcut}
\eeq
where (b) and (c) denote the gluon bremsstrahlung- and the
quark-pair emission process, respectively, and
$\Gamma^{0}_{77}$ is given in Eq.~(\ref{gammadecomp}).
Like in Section~\ref{sec:o7calc} we use
a regulator mass $m_f$ for the secondary quark-antiquark pair which means that
Eq.~(\ref{eq:gamcut}) 
can be calculated in $d=4$ dimensions and with $m_s=0$.

The calculation for the gluon bremsstrahlung piece  
$\hat{\Gamma}_{77}^{(2),(b)}(E_\gamma \leq E_{\rm cut})$ is straightforward.
Adopting the notation 
\beq
  E_{\rm cut} = \frac{m_b}{2} (1-\d) = E_{\rm max} (1-\d)\, ,
\eeq
the result reads
\begin{eqnarray}
  \hat{\Gamma}_{77}^{(2),(b)}(E_{\g} \leq E_{\rm cut}) &=& 
  \left(\frac{\a_s}{4\p}\right)^2 \frac{4 C_F T
  n_f}{9} \left(31 -30\d -3\d^2 +2\d^3 +21\ln(\d) \right. \nn \\
  && \left. +12\d\ln(\d) -3\d^2\ln(\d) +6\ln^2(\d)\right) 
  \left(\ln(f)+2\mub\right),
\end{eqnarray}
with $f=m_f^2/m_b^2$.

The calculation for $\hat{\Gamma}_{77}^{(2),(c)}(E_{\g} \leq E_{\rm cut})$ is
somewhat more involved but still can be performed analytically,
yielding
\begin{eqnarray}
  \label{eq:cresult}
  \hat{\Gamma}_{77}^{(2),(c)}(E_{\gamma}\leq E_{\text{cut}}) &=&
  \left(\alphasFourPi\right)^2 \frac{2C_F T n_f}{9}\Bigg(-147 -9\pi^2
  +48\zeta(3)-48\text{Li}_3(\d) +54\text{Li}_2(\d) \nn\\ && -\ln(\d)\left(85
  -4\pi^2 -54\ln(1-\d) -24\text{Li}_2(\d)\right) +13\ln^2(\d) +12\ln^3(\d)
  \nn\\ && +\d \Big(160 +4\pi^2 -24\text{Li}_2(\d) -24\ln(1-\d)\ln(\d)
  -94\ln(\d) +36\ln^2(\d)\Big) \nn\\ && +\d^2\bigg(1-\pi^2 +6\text{Li}_2(\d)
  +6\ln(1-\d)\ln(\d) +19\ln(\d) -9\ln^2(\d)\bigg) \nn\\ && -\d^3\bigg(14
  -4\ln(\d)\bigg) -2\ln(f)\bigg(31 -30\d -3\d^2 +2\d^3 +21\ln(\d)\nn\\ &&
  +12\d\ln(\d) -3\d^2\ln(\d)+6\ln^2(\d)\bigg)\Bigg).
\end{eqnarray}
Note that the sum of
$\hat{\Gamma}_{77}^{(2),(b)}(E_{\g} \leq E_{\rm cut})$ and
$\hat{\Gamma}_{77}^{(2),(c)}(E_{\g} \leq E_{\rm cut})$ is finite in the limit
$m_f \to 0$. This completes the calculation of 
$\Gamma^{(2),n_f}_{77}(b \to X_s \gamma)_{E_\gamma \leq E_{\rm cut}}$, defined
in \Eref{eq:gamcut}.

We note that differentiating
$\Gamma_{77}^{(2),n_f}(b \to X_s \gamma)_{E_\gamma
\le E_{\rm cut}}$ with respect to the photon energy cut $E_{\rm cut}$
generates the corresponding term of order $\alpha_s^2n_f$ 
to the photon energy spectrum. The result we obtain is in complete agreement 
with Eq.~(9) of Ref.~\cite{Ligeti:1999ea}, where ${\cal O}(\alpha_s^2n_f)$
corrections to the photon energy spectrum were calculated.




\begin{thebibliography}{99}

\bibitem{Chen:2001fj}
S.~Chen {\it et al.}  (CLEO Collaboration),
Phys.\ Rev.\ Lett.\  {\bf 87} (2001) 251807.

\bibitem{Abe:2001hk}
K.~Abe {\it et al.}  (Belle Collaboration),
Phys.\ Lett.\ B {\bf 511} (2001) 151.

\bibitem{Barate:1998vz}
R.~Barate {\it et al.}  (ALEPH Collaboration),
Phys.\ Lett.\ B {\bf 429} (1998) 169.

\bibitem{Aubert1}
B.~Aubert {\it et al.}  (BABAR Collaboration),
hep-ex/0207074.

\bibitem{Aubert2}
B.~Aubert {\it et al.}  (BABAR Collaboration),
hep-ex/0207076.

\bibitem{Jessop}
C.~Jessop, SLAC-PUB-9610.

\bibitem{Czarnecki:1998tn}
A.~Czarnecki and W.~J.~Marciano,
Phys.\ Rev.\ Lett.\  {\bf 81} (1998) 277.

\bibitem{Kagan:1999ym}
A.~L.~Kagan and M.~Neubert,
Eur.\ Phys.\ J.\ C {\bf 7} (1999) 5.

\bibitem{Baranowski:1999tq}
K.~Baranowski and M.~Misiak,
Phys.\ Lett.\ B {\bf 483} (2000) 410.

\bibitem{GH00}
P.~Gambino and U.~Haisch,
JHEP {\bf 0009} (2000) 001;
\\
P.~Gambino and U.~Haisch,
JHEP {\bf 0110} (2001) 020.

\bibitem{Hurth:2003vb}
T.~Hurth,
hep-ph/0212304.

\bibitem{Chetyrkin:1996vx}
K.G.~Chetyrkin, M.~Misiak, and M.~M\"unz,
Phys.\ Lett.\ B {\bf 400} (1997) 206
[Erratum-ibid.\ B {\bf 425} (1998) 414].

\bibitem{Ciuchini:1997xe}
M.~Ciuchini, G.~Degrassi, P.~Gambino, and G.~F.~Giudice,
Nucl.\ Phys.\ B {\bf 527} (1998) 21.

\bibitem{BG98}
F.~M.~Borzumati and C.~Greub,
Phys.\ Rev.\ D {\bf 58} (1998) 074004;
Phys.\ Rev.\ D {\bf 59} (1998) 057501.

\bibitem{Buras:1997}
A.~J.~Buras, A.~Kwiatkowski, and N.~Pott,
Nucl.\ Phys.\ B {\bf 517} (1998) 353; \\
A.~J.~Buras, A.~Kwiatkowski, and N.~Pott,
Phys.\ Lett.\ B {\bf 414} (1997) 157
[Erratum-ibid.\ B {\bf 434} (1998) 459].

\bibitem{Gambino:2001ew}
P.~Gambino and M.~Misiak,
Nucl.\ Phys.\ B {\bf 611} (2001) 338.

\bibitem{Buras:2002tp}
A.~J.~Buras, A.~Czarnecki, M.~Misiak, and J.~Urban,
Nucl.\ Phys.\ B {\bf 631} (2002) 219.

\bibitem{Ali:2002jg}
A.~Ali, E.~Lunghi, C.~Greub, and G.~Hiller,
Phys.\ Rev.\ D {\bf 66} (2002) 034002.

\bibitem{Buras:xp}
A.~J.~Buras, M.~Misiak, M.~M\"unz, and S.~Pokorski,
Nucl.\ Phys.\ B {\bf 424} (1994) 374.

\bibitem{Greub:1996tg}
C. Greub, T. Hurth, and D. Wyler,
Phys. Rev. D {\bf 54} (1996) 3350.

\bibitem{Beneke:1994qe}
M.~Beneke and V.~M.~Braun,
Phys.\ Lett.\ B {\bf 348} (1995) 513.

\bibitem{Brodsky:1982gc}
S.~J.~Brodsky, G.~P.~Lepage, and P.~B.~Mackenzie,
Phys.\ Rev.\ D {\bf 28} (1983) 228.

\bibitem{Luke:1994yc}
M.~E.~Luke, M.~J.~Savage, and M.~B.~Wise,
Phys.\ Lett.\ B {\bf 345} (1995) 301.

\bibitem{Czarnecki:1998kt}
A.~Czarnecki and K.~Melnikov,
Phys.\ Rev.\ D {\bf 59} (1999) 014036.

\bibitem{Ligeti:1999ea}
Z.~Ligeti, M.~E.~Luke, A.~V.~Manohar, and M.~B.~Wise,
Phys.\ Rev.\ D {\bf 60} (1999) 034019.

\bibitem{Greub:2000an}
C. Greub and P. Liniger,
Phys. Lett. B {\bf 494} (2000) 237.

\bibitem{Asatryan:2001zw}
H.~H. Asatryan, H.~M. Asatrian, C.~Greub, and M.~Walker.
Phys. Rev. D {\bf 65} (2002) 074004.

\bibitem{MMpriv}
M.~Misiak, private communication.

\bibitem{powermb}
A. Falk, M. Luke, and M. Savage, Phys. Rev. D {\bf 49} (1994) 3367; \\
I.I. Bigi, M. Shifman, N.G. Uraltsev, and A.I. Vainshtein, Phys. Rev. Lett.
 {\bf 71} (1993) 496; \\
A.V. Manohar and M.B. Wise, Phys. Rev. D {\bf 49} (1994) 1310;        \\
A. Falk, M. Luke, and M. Savage, Phys. Rev. D {\bf 53} (1996) 2491.

\bibitem{powermc}
M.B. Voloshin, 
Phys. Lett. B {\bf 397} (1997) 275;                     \\
Z. Ligeti, L. Randall, and M.B. Wise, Phys. Lett. B {\bf 402} 
(1997) 178 ;\\
A.K. Grant, A.G. Morgan, S. Nussinov, and R.D. Peccei, Phys. Rev. 
 D {\bf 56} (1997) 3151;                                           \\ 
G. Buchalla, G. Isidori, and S.J. Rey, Nucl. Phys. B {\bf 511} (1998)
594.

\bibitem{AG}
A. Ali and C. Greub, Zeitschr. \ Phys. \ C {\bf 49} (1991) 431; 
Phys.\ Lett.\ B {\bf 259} (1991) 182;
Phys.\ Lett.\ B {\bf 361} (1995) 146.

\bibitem{POTT}
N. Pott, Phys.\ Rev.\ D {\bf 54} (1996) 938.

\bibitem{Nir}
Y. Nir, Phys.\ Lett.\ B {\bf 221} (1989) 184.

\bibitem{Simma:1990nr}
H.~Simma and D.~Wyler,
Nucl. Phys. B {\bf 344} (1990) 283.

\end{thebibliography}
\end{document}